\documentclass[fleqn,usenatbib]{mnras}

\usepackage{newtxtext,newtxmath}
\usepackage{booktabs}

\usepackage[T1]{fontenc}
\usepackage{tabularray}
\DeclareRobustCommand{\VAN}[3]{#2}
\let\VANthebibliography\thebibliography
\def\thebibliography{\DeclareRobustCommand{\VAN}[3]{##3}\VANthebibliography}

\usepackage{graphicx}	
\usepackage{amsmath}	
\newcommand{\degree}{^\circ}

\title[Runaway OB stars within 1 kpc of the Sun]{Runaway OB stars within 1 kpc of the Sun}

\author[Juan Mart\'inez Garc\'ia]{
Juan Mart\'inez Garc\'ia$^{1}$\thanks{E-mail: j.martinez.garcia@keele.ac.uk},
Nicholas J. Wright$^{1}$ and 
Alexis L. Quintana$^{1,2,3}$
\\
$^{1}$Astrophysics Group, Keele University, Keele ST5 5BG, UK\\ 
$^{2}$Departamento de Física Aplicada, Facultad de Ciencias, Universidad de Alicante, Carretera de San Vicente s/n, 03690 San Vicente del Raspeig, Spain\\ 
$^{3}$LIRA, Observatoire de Paris, Université PSL, Sorbonne Université, Université Paris Cité, CY Cergy Paris Université, CNRS, 92190 Meudon, France
}

\date{Accepted XXX. Received YYY; in original form ZZZ}

\pubyear{\the\year{}}
\defcitealias{Quintana25}{Q25}
\begin{document}
\label{firstpage}
\pagerange{\pageref{firstpage}--\pageref{lastpage}}
\maketitle

\begin{abstract}
Runaway stars are high-velocity stars ejected from their birth environments that can provide insights into the kinematic history of the stellar cluster they were ejected from. We derived runaway star probabilities for 40 O-type stars and $24,487$ B-type stars taken from a recently published volume-complete sample of OB stars within 1 kpc of the Sun. We fit a Galactic rotation model to the observed proper motions of these stars and identify runaway stars using both a fixed 2D peculiar velocity threshold of $23\,km\,s^{-1}$ and by comparing individual peculiar velocities to the dispersion of the whole sample. We find runaway fractions of $17.5^{+0.1}_{-2.5}\%$ for O-type stars and $6.9\pm0.1\%$ for B-type stars; both using the fixed velocity threshold method. These values are consistent with previous studies, but with differences that are largely attributable to the underlying samples of OB stars used in various studies to identify runaway stars and to variations in the methods used to select them. 
\end{abstract}




\section{Introduction}

Runaway stars are stars that have been ejected from their birth environment. Traditionally, the term has been applied to OB stars, due to the ease of observing such objects at great distances, but can now be applied to any type of star. They are often isolated from other young stars and have large peculiar velocities \citep[e.g.][]{Blaauw1961}. 

\cite{Blaauw1961} observed that around $21\%$ of O-type stars have high velocities, and since then many authors have attempted to estimate the fraction of O-type stars that are runaway stars, commonly referred to as the runaway fraction, producing a wide dispersion of results. For example, \citet{Conti1977} found a fraction of $7\%$, \citet{Stone1979} found $54.7\%$, \cite{MaizApellaniz2018} found a fraction of $5.7\%$, while \cite{Carretero-Castrillo2023} found $23.7\%-25.1\%$ depending on the spectral subtype studied. This lack of consistency between results can mainly be attributed to different observational definitions of runaway stars and the underlying samples studied. 

B-type stars consistently show a lower runaway fraction than O-type stars. \cite{Blaauw1961} found that only $1.5-2.5\%$ of the B stars he studied were runaway stars, \cite{MaizApellaniz2018} found a fraction of $3.1\%$ and \cite{Carretero-Castrillo2023} found a fraction that goes from $6.2\%$ for early B-type stars to $4.8\%$ for later B-type stars.

The two main ejection mechanisms that have been suggested are the Binary Supernova Scenario \citep[BSS, e.g.][]{Blaauw1961, Portegies2000, Renzo2019} and the Dynamical Ejection Scenario \citep[DES, e.g.][]{Poveda1967, Leonard1990,Oh2016}. 

In the BSS, runaways are created after one of the stars in a binary system explodes as a supernova. The mass loss experienced by the star plus the conservation of momentum in an asymmetric explosion \citep[see][]{Renzo2019} disrupts the binary system and releases the companion star, often at a high velocity.

The DES consists of scattering produced by close encounters of two pairs of binary systems or a single star with a binary. The number of runaway stars and their velocities created from the DES depend on the density of the cluster and the properties of the binaries within the cluster \citep[][]{Perets2012, Oh2016}, meaning that these properties at the time of ejection can be constrained from studies of the number and velocity of runaway stars. Runaway stars can also provide stellar model-independent age estimates for clusters, as ejections peak when the clusters are between $0.5$ and $1.5$ Myr old \citep{Oh2016}. Furthermore, the ejection time of a star (the time required for it to travel from its birthplace to its current location) cannot be older than the age of the star, so runaway stars can provide a powerful constraint on stellar ages.

Determining which ejection mechanism produced a runaway star is crucial to be able to infer properties of the birth cluster. Many theoretical works have modelled the properties of runaway stars ejected by the DES \citep[eg.][]{Poveda1967, Leonard1990, Perets2012, Oh2016} and the BSS \citep[eg.][]{Portegies2000, Renzo2019}. The BSS mechanism is unlikely to produce binary runaway stars and that ejections produced by the DES mechanism should be more energetic than those from the BSS mechanism.  \cite{Phillips2024} showed that, in the Small Magellanic Cloud, OB stars are more likely to be ejected via the DES, except for stars that show emission features.

Studies that aim to find runaway stars can be divided between those that are targeted, with the goal of identifying the runaway stars from a certain cluster \citep[e.g.][]{ Drew2019, Schoettler2022,Stoop2023}, or untargeted, with the goal of identifying any runaway stars across the sky or in a large region of it \citep[e.g.][]{MaizApellaniz2018,Carretero-Castrillo2023, Guo2024}. Targeted studies are able to employ a more precise observational definition of a runaway star, since the distance to the cluster and the relative velocity of the runaway star can be accurately determined. The wider search area of untargeted studies allows them to find older runaway stars that may be further away from the cluster. Untargeted studies benefit from large populations of OB stars to obtain more accurate statistics, but are typically only possible with large scale surveys such as \textit{Gaia}.


In this paper we search for runaway OB stars within a recently-published volume-complete sample of OB stars within 1 kpc of the Sun, employing multiple methods to identify runaway stars and facilitating broad comparisons with the literature. In Section \ref{sec:2} we describe the sample of OB stars used. Section \ref{sec:3} is dedicated to the Galactic rotation model and the calculation of residual velocities for all stars. In Sections \ref{sec:4} and \ref{sec:literature_comparison} we identify runaway stars and discuss the results, before concluding in Section \ref{conclusions}.




\section{The sample of OB stars}
\label{sec:2}

In this section we describe the data used for this work, which is mostly taken from the catalogue of OB stars within 1 kpc of the Sun presented by \cite{Quintana25}.

\subsection{OB stars within 1 kpc}

\citet[hereafter Q25]{Quintana25} created an astro-photometric catalogue of OB stars within 1 kpc of the Sun based on an SED fitting tool first described in \cite{Quintana21} and \cite{Quintana23}. The SED fitting tool uses \textit{Gaia} astrometry combined with photometry from \textit{Gaia} and other optical and near-IR surveys. The observed SEDs are fitted to evolutionary and atmospheric stellar models, allowing stellar masses, ages, effective temperatures, luminosities and distances to be constrained. \textit{Gaia} astrometry is supplemented by astrometry from HIPPARCOS for the brightest stars that are saturated in \textit{Gaia}.

The catalogue contains $24,706$ OB stars within 1 kpc of the Sun, defined as stars with an SED-fitted median effective temperature greater than $10,000$ K. The catalogue is estimated to be over $95\%$ complete for OB stars within that volume. One of the advantages of \citetalias{Quintana25} is that this high completeness extends to late B-type stars, which are often overlooked in the study of runaway stars.

\subsection{Spectral type classification}

We refine the sample of OB stars from \citetalias{Quintana25} using spectral types from the literature. This was particularly important for the hottest stars, for which it was important to establish a reliable list of O-type stars within our survey volume. We used the SIMBAD database to search for spectral types for our sources and where these were of sufficient quality (A, B or C) we replaced our SED-fitted effective temperatures with spectroscopic effective temperatures.

To identify any O-type stars within our survey volume that might have been missed by \citetalias{Quintana25} we searched the Galactic O-type Star Catalogue \citep[GOSC, ][]{GOSC}, which is the most up-to-date list of O-type stars in the Milky Way. We also searched the catalogue of \cite{Garmany82}, which includes a list of Galactic O-stars with high completeness up to 2.5 kpc of the Sun, but did not find any O-type star not already included in GOSC. 

 Our final list of O-type stars, shown in Table \ref{tab:True_True}, contains 48 sources and is divided into three subsets: (1) Stars that are present in both GOSC and \citetalias{Quintana25} (36 objects), (2) Stars in GOSC with an astrometric distance within 1 kpc that are not present in \citetalias{Quintana25} (9 objects) and (3) stars with high SED-fitted effective temperature (over $30,463$ K) but without a spectral type in the literature (4 objects). Additionally, we include a list of stars that we have not included in our list of O-type stars within 1 kpc of the Sun (Table \ref{tab:rejected}). Stars that are spectroscopically confirmed O-type stars are removed from the sample of B-type stars.





\begin{table*}
\caption{Final list of O-type stars within 1 kpc of the Sun. First part: Spectroscopic O-type stars from GOSC that are included in \citetalias{Quintana25} Second part: Spectroscopic O-type stars from GOSC not present in \citetalias{Quintana25} . Third part: stars with a SED fitted temperature over $30,463$ K (equivalent to spectral type O), but unknown spectral type. $D_{XY}$ is the distance projected on the Galactic plane, from \citetalias{Quintana25} when available, \citet{Bailer-Jones2021} (\dag) otherwise. In cases where none of the above were available we used the HIPPARCOS parallax (\ddag) to calculate the distance.  Spectral type (GOSC SpT) and luminosity class (LC) references: S11a: \citet{S11a}; S14: \citet{S14};  M16a: \citet{M16a}; M18a: \citet{M18a}; M18b: \citet{M18b}; M19: \citet{M19a}. Last column tracks whether the star is associated to a cluster in GOSC (y: yes; n: no).}
\renewcommand{\arraystretch}{1.2} 

\begin{tabular}{llllll}
\toprule
HDMHDE & STARS & $D_{XY} (pc)$ & Spectral Type & REFV3 & $\log T_{\rm eff}/K$ \\
\midrule
149 757 & zeta Oph                & $113_{-4}^{+5}$ & O9.2IV & S14  & $4.23_{-0.05}^{+0.12}$ \\
154 643 & HD 154 643              & $936_{-45}^{+59}$ & O9.7III & S14  & $4.34_{-0.02}^{+0.09}$ \\
34 656  & HD 34 656               & $995_{-108}^{+87}$ & O7.5II & S11a & $4.34_{-0.14}^{+0.11}$ \\
216 532 & HD 216 532              & $775_{-14}^{+14}$ & O8.5V & S11a & $4.35_{-0.05}^{+0.09}$ \\
165 174 & HD 165 174              & $974_{-121}^{+172}$ & O9.7II & S14  & $4.36_{-0.03}^{+0.12}$ \\
47 839  & 15 Mon AaAb             & $326_{-53}^{+92}$ & O7V & M18a & $4.37_{-0.09}^{+0.05}$ \\
93 521  & HD 93 521               & $1606_{-475}^{+1308}$ & O9.5III & S11a & $4.40_{-0.07}^{+0.12}$ \\
209 339 & HD 209 339              & $968_{-69}^{+95}$ & O9.7IV & S14  & $4.40_{-0.06}^{+0.08}$ \\
37 743  & zeta Ori B              & $296_{-63}^{+98}$ & O9.7III & M18a & $4.41_{-0.09}^{+0.07}$ \\
206 183 & HD 206 183              & $912_{-41}^{+41}$ & O9.5IV-V & S11a & $4.41_{-0.05}^{+0.04}$ \\
207 538 & HD 207 538              & $838_{-32}^{+37}$ & O9.7IV & S11a & $4.41_{-0.04}^{+0.04}$ \\
45 314  & PZ Gem                  & $929_{-97}^{+115}$ & O9: & S11a & $4.42_{-0.48}^{+0.10}$ \\
24 534  & X Per                   & $608_{-41}^{+53}$ & O9.5: & S11a & $4.42_{-0.07}^{+0.05}$ \\
214 680 & 10 Lac                  & $513_{-86}^{+318}$ & O9V & S11a & $4.43_{-0.04}^{+0.07}$ \\
36 861  & lambda Ori A            & $316_{-61}^{+176}$ & O8III & S11a & $4.43_{-0.10}^{+0.10}$ \\
204 827 & HD 204 827 AaAb         & $981_{-74}^{+52}$ & O9.5IV & S14  & $4.44_{-0.10}^{+0.08}$ \\
36 512  & upsilon Ori             & $423_{-41}^{+77}$ & O9.7V & S11a & $4.44_{-0.03}^{+0.03}$ \\
38 666  & mu Col                  & $663_{-95}^{+272}$ & O9.5V & S14  & $4.45_{-0.04}^{+0.04}$ \\
57 060  & 29 CMa                  & $654_{-127}^{+215}$ & O7Ia & S14  & $4.45_{-0.12}^{+0.10}$ \\
216 898 & HD 216 898              & $824_{-10}^{+9}$ & O9V & S11a & $4.45_{-0.07}^{+0.04}$ \\
36 486  & delta Ori Aa            & $225_{-27}^{+39}$ & O9.5II & M18a & $4.47_{-0.09}^{+0.05}$ \\
37 742  & zeta Ori AaAb           & $250_{-44}^{+89}$ & O9.2Ib & M18a & $4.47_{-0.09}^{+0.11}$ \\
37 041  & theta\^2 Ori A           & $432_{-4}^{+4}$ & O9.2V & M19a & $4.48_{-0.08}^{+0.04}$ \\
217 086 & HD 217 086              & $871_{-43}^{+40}$ & O7V & S14  & $4.48_{-0.10}^{+0.06}$ \\
75 759  & HD 75 759 AB            & $993_{-38}^{+105}$ & O9V & S14  & $4.50_{-0.09}^{+0.04}$ \\
37 022  & theta\^1 Ori CaCb        & $440_{-10}^{+30}$ & O7 & M19a & $4.50_{-0.10}^{+0.07}$ \\
199 579 & HD 199 579              & $844_{-14}^{+14}$ & O6.5V & S11a & $4.51_{-0.09}^{+0.05}$ \\
24 431  & HD 24 431 A             & $997_{-99}^{+152}$ & O9III & S11a & $4.51_{-0.11}^{+0.07}$ \\
37 468  & sigma Ori AaAb          & $456_{-214}^{+721}$ & O9.5V & M19a & $4.51_{-0.27}^{+0.10}$ \\
66 811  & zeta Pup                & $341_{-20}^{+38}$ & O4I & S14  & $4.55_{-0.11}^{+0.06}$ \\
28 446A & 1 Cam A                 & $798_{-36}^{+132}$ & O9.7II & S14  & $4.55_{-0.09}^{+0.05}$ \\
135 591 & HD 135 591              & $937_{-43}^{+143}$ & O8IV & S14  & $4.56_{-0.09}^{+0.05}$ \\
37 043  & iota Ori AaAb           & $407_{-24}^{+7}$ & O8.5III & M19a & $4.57_{-0.09}^{+0.08}$ \\
24 912  & xi Per                  & $907_{-233}^{+1049}$ & O7.5III & S11a & $4.57_{-0.09}^{+0.09}$ \\
203 064 & 68 Cyg                  & $892_{-36}^{+106}$ & O7.5III & S11a & $4.58_{-0.12}^{+0.07}$ \\
210 839 & lambda Cep              & $941_{-90}^{+175}$ & O6.5I & S11a & $4.59_{-0.11}^{+0.08}$ \\
\midrule
34 078 & AE Aur &$382_{-5}^{+5}$\dag &O9.5V&S11a& \\
73 882 & NX Vel AB & $461 \pm 193$\ddag & O8.5IV& S14& \\
111 886 & HD 114 886 AaAb&$469 \pm 205$\ddag & O9III& S14& \\
135 204 & Delta circ AaAbAc&$921_{-284}^{+347}$\dag &O7IV&M16a& \\
145 217 & HD 145 217   &$ 752^{+113}_{-57}$\dag &O8 V      & M16a   & \\
155 889 & HD 155 889 AB & $832 \pm 395$\ddag & O9.5IV & S14& \\
159 176 & HD 159 176 &$790_{-112}^{+172}$\dag &O7V&S14& \\
190 429 & HD 190 429B &$ 861 \pm 468$ \ddag & O9.5II-III& S11a& \\
206 267 & HD 206 267 AaAb& $704_{-53}^{+95}$\dag &O6V&M19A& \\
\midrule
 169753 & V* RZ Sct &$817^{+77}_{-63}$ & & & $4.51_{-0.11}^{+0.06}$\\
74804 & HD 74804 & $973^{+21}_{-18}$ & & & $4.50_{-0.11}^{+0.06}$  \\
219634  & V* V649 Cas & $868^{+211}_{-16}$ & & & $4.52_{-0.09}^{+0.05}$\\

\bottomrule
\end{tabular}

\label{tab:True_True}

\end{table*}

Finally, we exclude stars that only have HIPPARCOS proper motions, as the larger uncertainties of Hipparcos data compared to that of Gaia data makes these objects very unreliable for our analysis.

The final sample contains $24,488$ B-type stars and $40$ O-type stars. 

\section{Galactic Rotation model}
\label{sec:3}

In this section we describe the model for Galactic rotation and show the results obtained after fitting the model to the sample described in Section 2. We follow a similar approach to \cite{Almannaei2024}, which consists of using a Markov Chain Monte Carlo (MCMC) method to fit the average motion of the stars in our sample to a model for the Galactic rotation. We then fit a Maxwell-Boltzmann distribution to the residual (Galactic model subtracted) velocities.

The first step is to convert the positions ($\alpha$, $\delta$, $\varpi$) and proper motions ($\mu_\alpha^*=\mu_\alpha cos(\delta)$, $\mu_\delta$) to a Galactic coordinate system ($\ell$, $b$, $r$, $V_\ell^{helio}$, $V_b$), where $\ell$ and $b$ are the Galactic coordinates, $r$ is the astro-photometric distance derived by \citetalias{Quintana25}, and $V_\ell^{helio}$ and $V_b$ are the components of the transverse velocity of the star in the $l$ and $b$ directions, respectively. The conversion of the quantities and their associated uncertainties was made following the equations laid out in \cite{GaiaDocumentation}.

\begin{center}
    \begin{figure}
        \centering
        \includegraphics[width=\linewidth]{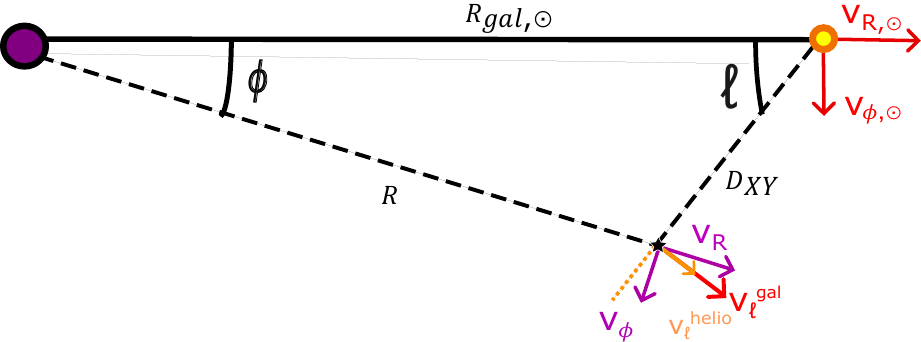}
        \caption{Diagram of the coordinate system used in the XY-plane of the Galaxy. The large purple circle represents the Galactic Centre, while the small yellow circle represents the Sun. The star represents any given star. The arrows show the direction of different velocity components in the Galactic plane.}
        \label{fig:coord_diagram}
    \end{figure}
\end{center}

Additionally, we use the reference system (R, $\phi$) shown in Figure \ref{fig:coord_diagram}, where $R$ is the distance from the Galactic Centre to a given star and is calculated as $\sqrt{(R_{gal,\odot}-X)^2+Y^2}$, with $R_{gal,\odot}$ being the distance of the Sun to the Galactic Centre, and $X$ and $Y$ being the Galactic Cartesian coordinates of the star with respect to the Sun. $\phi$ is the angle between the line that joins the Galactic Centre with the Sun and the line that joins the Galactic Centre with the star, and is calculated as $\phi=\arctan{\dfrac{Y}{R_{gal,\odot}-X}}$. We use the value $R_{gal,\odot}=8178\pm 25$ pc \citep{GRAVITY2019}. It is also useful to define the distance of the star projected on the Galactic plane, $D_{XY}=r\cos{b}$.

The second step is to create and fit a model of Galactic rotation. We assume that $V_b$ is not dependent on sky position (see Figure \ref{fig:peculiar velocities} for verification of this assumption), and therefore the peculiar velocity in this direction can be modelled as $\Tilde{V_b} =V_b-\Bar{V_b}$, where $\Bar{V_b}$ is the mean velocity in the $b$ direction.

To model the motion in the $\ell$ direction, we start by calculating the galactocentric velocity, $V_l^{gal}$, by transforming the velocities measured in the reference frame of the Sun ($V_\ell^{helio}$) to the reference frame of the Galactic Center as follows (see Figure \ref{fig:coord_diagram}):
\begin{equation}
V_\ell^{gal}=V_\ell^{helio}+V_{R,\odot}\sin(\ell)+V_{\phi,\odot}\cos(\ell).
\end{equation}

\noindent where $V_{R,\odot}$ and $V_{\phi,\odot}$ are respectively the velocity of the Sun in the direction away from the Galactic Centre and azimuthally. The values of these quantities are $V_{R,\odot}=-10\pm1 \ km\,s^{-1}$ \citep{Bland2016} and $V_{\phi,\odot}=247.4\pm 1.4 \,km\,s^{-1}$ \citep{GRAVITY2019}. Note that the value of $V_{R,\odot}$ is negative in the coordinate system defined in Figure \ref{fig:coord_diagram}, as the Sun's Galactic radial velocity is positive towards the Galactic Centre. Using these values, we can calculate $V_l^{gal}$ for all our sources.

The motion of the star in the Galactic plane can also be written as the combination of a rotational component $V_\phi$ and a radial component $V_R$. Projecting each component onto the $l$ direction we can calculate the galactocentric velocity (see Figure \ref{fig:coord_diagram}):
\begin{equation}
    V_\ell^{gal}= V_\phi \cos(\phi+\ell)+V_R\sin(\phi+\ell)
    \label{projection}
\end{equation}
In general, the velocities in the $\phi$ and $R$ directions are an unknown function of the Galactic radius $R$. We can truncate the Taylor expansion of both components to first order, yielding:
\begin{equation}
    \begin{array}{cc}
         &  V_\phi\simeq V_{\phi,0}+ \dfrac{dV_\phi}{dR}\Big{|}_0(R-R_{gal,\odot}) \\
         \\
         & V_R\simeq V_{R,0}+ \dfrac{dV_R}{dR}\Big{|}_0(R-R_{gal,\odot})
    \end{array}
\end{equation}

Applying this Taylor expansion to Eq. \ref{projection} we get: 

\begin{multline}
    V_\ell^{gal}\simeq \left(V_{\phi,0} + \dfrac{dV_\phi}{dR}\Big{|}_0(R-R_{gal,\odot})\right) \cos(\phi+\ell) \hspace{0.5cm} + \\ \left(V_{R,0}+ \dfrac{dV_R}{dR}\Big{|}_0(R-R_{gal,\odot})\right)\sin(\phi+\ell) \equiv V_\ell^{model}(R,\ell)
    \label{eq:model};
\end{multline}
\noindent where we define the model velocity $V_\ell^{model}(R,\ell)$.

For each star we know $\phi$, $\ell$, $R$ and its velocity in the $\ell$ direction $V^{gal}_\ell$. We can therefore model $V^{gal}_\ell$ using Eq. \ref{eq:model} treating $V_{\phi,0}$, $ \dfrac{dV_\phi}{dR}\Big{|}_0$, $V_{R,0}$ and $ \dfrac{dV_R}{dR}\Big{|}_0$ as free parameters. 

In the following subsections we describe the model-fitting process and the resulting best fit parameters.

\subsection{Fitting the model to the data}
\label{sec:fit}
Let $$v_\ell(\theta|R,\ell)=V_\ell^{gal}-V_\ell^{model}(\theta|R,\ell)=(V_{\ell,1},V_{\ell,2},...,V_{\ell,N})$$ be a vector containing the excess velocities ($V_\ell^{gal}$ are the observed velocities and $V_\ell^{model}(\theta|R,\ell)$ are the model velocities) for every star using a set of parameters $\theta=\{V_{\phi,0}, \dfrac{dV_\phi}{dR}\Big{|}_0,V_{R,0} ,  \dfrac{dV_R}{dR}\Big{|}_0\}$. Note that here we indicated explicitly the dependence of the model velocity on the model parameters.

We assume that the real excess velocities ($v'_\ell$, to distinguish them from the measured excess velocities $v_\ell$) are distributed according to a Gaussian with standard deviation $\sigma$, denoted as $G(v'_\ell)$.

Given the asymmetrical measurement uncertainty $(\Delta^+_{V,i},\ \Delta^-_{V,i})$, the probability that a value $v_\ell$ was measured given a real value $v_\ell'$ can be quantified as:


\begin{equation}
f_i(v_{\ell,i},v'_{\ell,i}) =
\begin{cases}
\displaystyle A \exp\left( -\frac{(v_{\ell,i}-v'_{\ell,i})^2}{2(\Delta^-_{V,i})^2} \right) & \text{if } v_{\ell,i}<v'_{\ell,i} \\
\displaystyle A \exp\left( -\frac{(v_{\ell,i}-v'_{\ell,i})^2}{2(\Delta^+_{V,i})^2} \right) & \text{if } v_{\ell,i} \geq v'_{\ell,i} 
\end{cases}
\label{eq:asy_norm}
\end{equation}

where $A=\frac{2}{\sqrt{2\pi(\Delta^+_{V,i}+ \Delta^-_{V,i})^2}}$.

Therefore, the probability that a value of $v_{\ell,i}$ is measured given a model with parameters $\theta$ is 

\begin{equation}
    P_i(v_{\ell,i};\theta)=\int_{-\infty}^{+\infty}f(v_{\ell,i},v'_{\ell,i})G(v'_{\ell,i})dv'_{\ell,i}
\end{equation}

The log-likelihood function for the complete set of data points will therefore be:
\begin{equation}
    L(v_{\ell};\theta)=\sum_i\log(P_i(v_{\ell,i};\theta))
\end{equation}

We want to maximise the log-likelihood to obtain the best fitting set of model parameters. To do this, we run an MCMC code with 100 walkers and 6000 steps with the likelihood function above, using the Python implementation {\it emcee} by \cite{MCMC}. We set our initial conditions for the model parameter to match the values obtained by \cite{Akhmetov2024} (see Table \ref{tab:best_fit}) plus a small random number for each walker. 

\begin{table*}
    \centering
    \caption{Comparison between our best fit parameters and values in the literature. Uncertainties of our parameters are based on the 16th and 84th percentiles from the posterior distribution of our MCMC simulation. The values from \citet{Akhmetov2024} have been adapted to match our definition of the azimuthal angle, which is positive in the opposite direction. The empty values mean that that parameter was assumed to be null by the model.}
    \begin{tabular}{l|l|c|c|c|c|c|c}
        \toprule
        Paper& Type & Distance & $V_{\phi,0} $&  
        $\dfrac{dV_\phi}{dR}\Big{|}_0$ & 
        $V_{R,0}$ &
        $\dfrac{dV_R}{dR}\Big{|}_0$ & $\sigma $\\ 
         & & & km s$^{-1}$  & km s$^{-1}$  kpc$^{-1}$& km s$^{-1}$ & km s$^{-1}$ kpc$^{-1}$ & km s$^{-1}$ \\
        \midrule
        \cite{Akhmetov2024}& All& $D_{XY}<1$ kpc & $232.19\pm 0.08$& $-1.00\pm 0.07$& $0.60\pm 0.09$ & $-5.11 \pm 0.09$ \\
        \cite{Almannaei2024}&OB&$D_{XY}<0.5$ kpc & $238.30\pm 2.43$& $1.82\pm 1.17$&  & & \\
        \cite{Almannaei2024}&OB& $D_{XY}<2$ kpc & $233.95\pm 2.24$& $-3.31\pm 0.48$&  & & \\
        This paper &OB &$D_{XY}<1kpc$ & $236.52\pm0.11$& $-0.73\pm 0.17$& $0.44\pm 0.10$ & $-3.70 \pm 0.29$ & $11.11\pm0.06$ \\
        \bottomrule
    \end{tabular}
    \label{tab:best_fit}
\end{table*}

\subsection{Best fit parameters and residual velocities}
Table \ref{tab:best_fit} lists the best fit parameters obtained in this paper and those obtained by \cite{Akhmetov2024} and \cite{Almannaei2024}. \cite{Akhmetov2024} selected all stars in \textit{Gaia} DR3 with six dimensional phase-space velocity, and fitted them to a model consisting of the velocities, and their first and second derivative in every direction. \cite{Almannaei2024} fitted OB stars within 2 kpc of the Sun to a model composed of the velocity in the $\ell$ direction, its first radial derivative and an additional term to model the non-axisymmetric components.  

The values of $V_{\phi,0}$ and $\dfrac{dV_\phi}{dR}\Big{|}_0$ that we have derived lie in between the values obtained by \cite{Almannaei2024} for stars with $D_{XY}<0.5$ and $D_{XY}<2kpc$, which is expected given that our sample is limited to stars with $D_{XY}<1kpc$, and both our sample and theirs are composed of OB stars. Our value of $\dfrac{dV_\phi}{dR}\Big{|}_0$ is consistent within the uncertainties with that of \cite{Akhmetov2024}, but there is a significant difference between our values and theirs for the rest of the parameters, likely caused by the fact that the kinematics of OB stars can differ from the kinematics of other stellar populations.

\begin{center}
    \begin{figure}
        \centering
        \includegraphics[width=\linewidth]{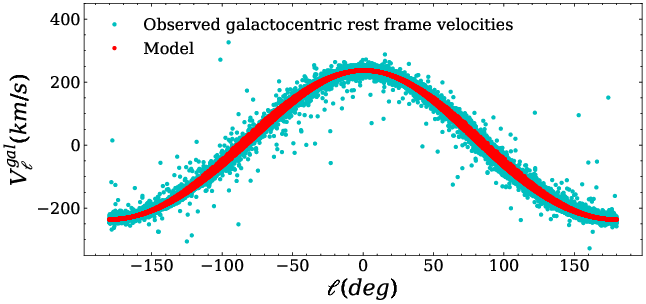}
        \caption{Observed galactocentric velocity in the $l$ direction (blue) and best fit model (red) for all the stars in our sample. }
        \label{fig:galactic_model}
    \end{figure}
\end{center}

Figure \ref{fig:galactic_model} shows the distribution of velocities in the $\ell$ direction and the best fitting model for each star. The model shows some thickness due to its dependence on parameters other than $\ell$, and since the model value for each star is shown rather than a line showing a model prediction at each value of $\ell$.

\begin{center}
    \begin{figure}
        \centering
        \includegraphics[width=\linewidth]{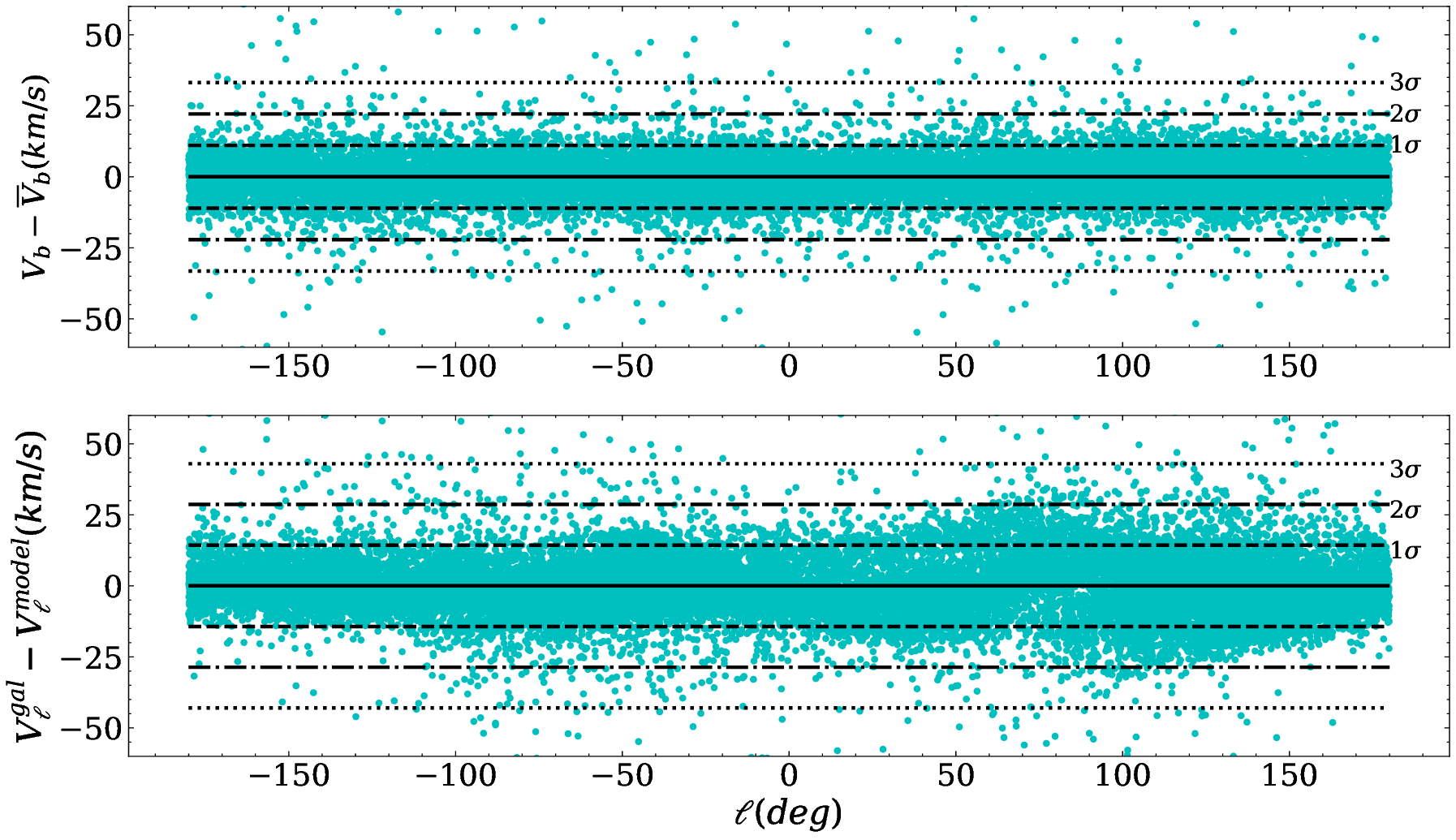}
        \caption{Peculiar velocities in the $b$ (top) and $\ell$ (bottom) directions. The solid lines represent the medians, which are by definition $0$ in the $b$ direction and $-0.2\,km\,s^{-1}$ in the $\ell$ direction. The dashed, dash-dotted and dotted lines represent $1$, $2$ and $3$ standard deviations from the median. The peculiar velocities in the $\ell$ direction are essentially the residuals from the Galactic rotation model.}
        \label{fig:peculiar velocities}
    \end{figure}
\end{center}

In Figure \ref{fig:peculiar velocities} the peculiar velocities are shown. In the $b$ direction, the peculiar velocities are independent of $\ell$. In the $\ell$ direction, although the median peculiar velocity is very small ($0.7\,km\,s^{-1}$), there is still some structure, mainly a small hike around $\ell=80\degree$, which might be caused by non-axisymmetric movements within the Milky Way. 


\subsection{Fitting a Maxwell-Boltzmann Distribution to the residual velocities}

The velocity distribution of a group of stars of equal mass interacting gravitationally is expected to follow a Maxwell-Boltzmann distribution. Characterizing this distribution will help us determine an appropriate threshold velocity, reducing the relative arbitrariness in the runaway star selection criteria (see Section \ref{sec:selection}). 

The Maxwell-Boltzmann distribution is such that the probability of a star having a velocity in a certain velocity range is 
\begin{equation}
    \int f(v)dv^3=\int \alpha e^{-v^2/b^2}dv^3,
    \label{eq:boltzmann}
\end{equation}
\noindent where $v$ is the velocity.

Using spherical coordinates, we can write $dv^3=v^2 \sin{\theta} \ dvd\phi d\theta$. Integrating over $\phi$ and $\theta$ in Eq. \ref{eq:boltzmann}, we get the distribution \begin{equation}
    \int 4\pi\alpha e^{-v^2/b^2}dv=\int f_v(v)dv
\end{equation}
Renaming $a=4\pi \alpha$ the distribution becomes:
\begin{equation}
    f_v(v)=av^2e^{-v^2/b^2}
    \label{eq:Boltzmann_spatial}
\end{equation}

If we now use cylindrical coordinates, $dv^3=v_\parallel dv_\parallel dv_\perp d\phi$, and $v^2=v_\parallel^2+v_\perp^2$. $v_\parallel$ corresponds to the projection of the velocities on a given plane, while $v_\perp$ is the projection on the axis perpendicular to said plane. Integrating over $v_\perp$ between $(-\infty,\infty)$ and over $\phi$ between $(0,2\pi)$ we get the distribution:
\begin{equation}
    \int 2\pi^{3/2}b \alpha v_\parallel e^{-v_\parallel^2/b^2}dv_\parallel=\int f_\parallel(v_\parallel)dv_\parallel
\end {equation}
Renaming the constant $\Tilde{a}=2\pi^{3/2}b \alpha=ab \sqrt{\pi}/2$ we arrive at the distribution:
\begin{equation}
    f_\parallel(v_\parallel)=\Tilde{a}v_\parallel e^{-v_\parallel^2/b^2}
    \label{eq:Boltzmann_parallel}
\end{equation}

\begin{center}
    \begin{figure*}
        \centering
        \includegraphics[width=\linewidth]{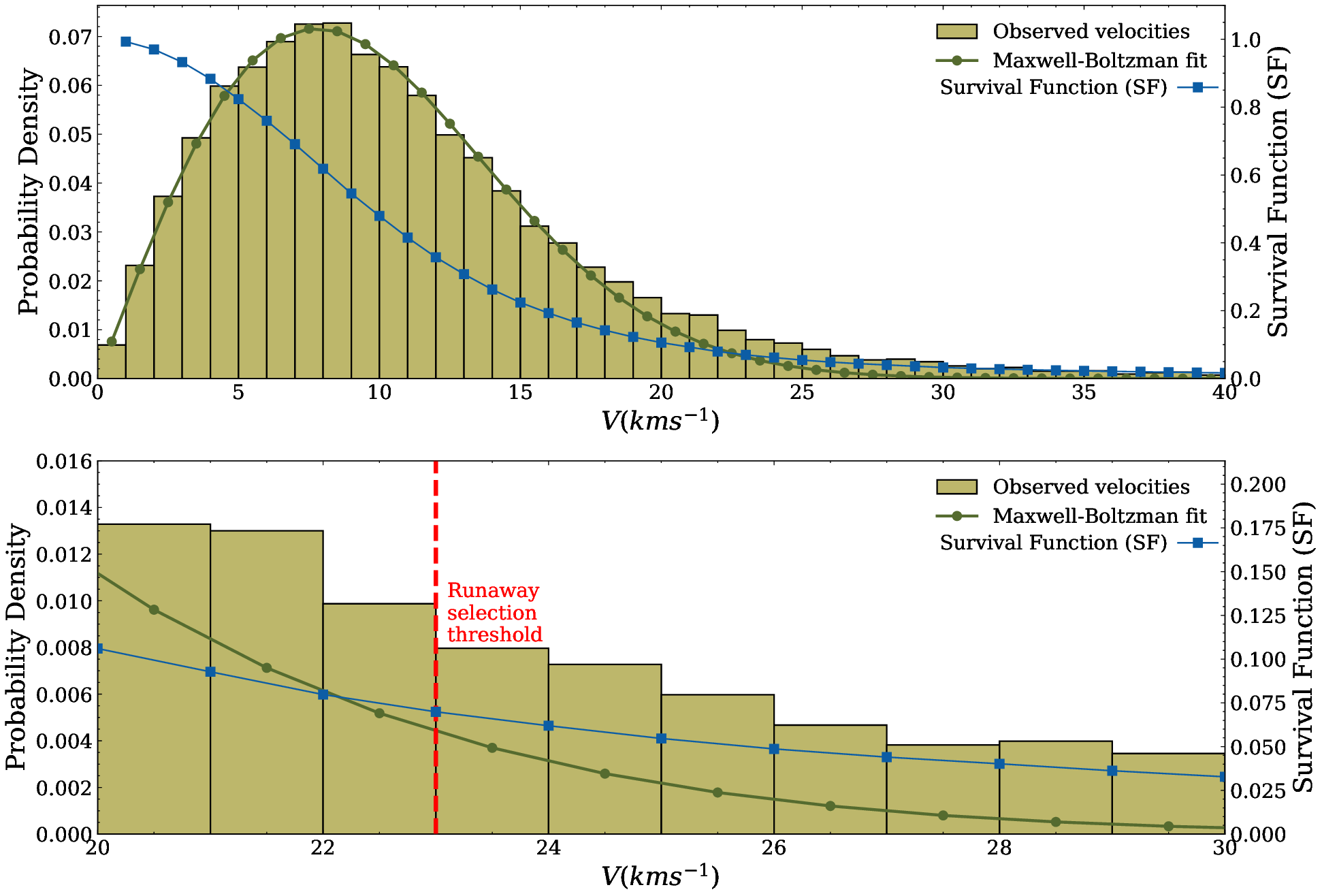}
        \caption{Histogram showing the distribution of 2D residual velocities defined as $V=\sqrt{(V_\ell^{gal}-V_\ell^{model})^2+(V_b-\overline{V}_b)^2}$. Also shown are the Maxwell-Boltzmann fit to this distribution, represented by the dark green line marked with circles, and the Survival Function (SF, defined as $1-$ {\it Cumulative Distribution Function}), represented by the blue line with square markers. Note that the figure uses two different Y-axes: the left Y-axis displays the probability density of peculiar plane velocities along with the Maxwell–Boltzmann fit, while the right Y-axis corresponds to the SF. The upper panel shows the full distribution, while the lower panel shows a close up of the region between 20 - 30 km s$^{-1}$.}
        \label{fig:histogram}
    \end{figure*}
\end{center}

Fitting the residual velocities to a Maxwell-Boltzmann distribution given by Eq. \ref{eq:Boltzmann_parallel} using {\tt scipy.optimize}, we determine the best-fit parameters $\Tilde{a}=0.0153\pm0.0002  \ (km\, s^{-1})^{-2}$ and $b=10.99 \pm0.06 \ km \, s^{-1}$. The residual velocity distribution and the Maxwell-Boltzmann fit are displayed in Figure \ref{fig:histogram}.

We can use the normalisation of the distribution as a test of how well it describes the data. We have not required normalisation as a requisite for the Maxwell-Boltzmann distribution function (Eqs. \ref{eq:boltzmann}, \ref{eq:Boltzmann_spatial} and \ref{eq:Boltzmann_parallel}). The normalisation of the probability distribution implies the following: 

\begin{equation}
    N=\dfrac{1}{2}\tilde{a}b^2=\dfrac{\sqrt{\pi}}{4}ab^3=1.
\end{equation}

For our best-fit parameters, the value we obtain is $N=0.92\pm 0.02$. Although it is close to 1, it is not compatible within the uncertainty, which shows that the distribution is not perfectly Maxwellian. This is expected for multiple reasons. Firstly, stars have different masses, and thermal equilibrium would imply that each star follows a distribution function where the dispersion is proportional to the inverse of the mass due to equipartition of energy. Furthermore, in our fit we are including stars that have been ejected, which do not form part of the distribution. Also, some OB stars, particularly the most massive ones, are younger than the timescale in which they reach thermal equilibrium (note that for an OB association with a diameter of $150 \ pc$ and a velocity dispersion of $4.5\, km \, s^{-1}$ the timescale for thermal equilibrium would be around $30 \ Myr$). 

\cite{Guo2024} fitted the spatial velocity of stars in their sample to a Maxwell-Boltzmann distribution given by Eq. \ref{eq:Boltzmann_spatial}. They fitted the histogram as opposed to the probability density distribution, meaning that their $a$ parameter has to be divided by the number of stars in their sample times the bin width of their histogram in order to normalise the area under the curve. The values obtained by \cite{Guo2024} are $\Tilde{a}_{Guo}=7.20\times 10^{-3}\pm 0.05\times 10^{-3}$;
$b_{Guo}=15.46\pm0.03\,km\,s^{-1}$, which are respectively smaller and larger than our values, and not consistent within the uncertainties. 

The mean velocity, the peak-likelihood velocity and the standard deviation of the velocity are all proportional to the value of $b$. This means that the stars in the sample from \cite{Guo2024} have on average higher velocities and a wider dispersion. This is because their Galactic rotation model only considers a constant rotation term, which in turn overestimates the velocities. 

\section{Runaway stars}

\label{sec:4}

Now that we have characterised the motion of the stars in the sample, we will use their peculiar velocities and the Maxwell-Boltzmann distribution to select and identify the runaway stars within 1 kpc of the Sun.

\subsection{Selection criteria}
\label{sec:selection}
We have selected runaway stars in two different ways:
\begin{itemize}
    \item {\bf Using a fixed peculiar velocity threshold:} We select $23\,km\,s^{-1}$ as the 2D velocity threshold, corresponding to the peculiar velocity at which stars are more likely than not to be a runaway star, as it can be seen on Figure \ref{fig:histogram} that the measured probability density is twice that predicted by the model for velocities higher than the threshold. The SF at a given velocity is equivalent to the fraction of stars that are runaway stars when that velocity is used as a threshold. The lower part of Figure \ref{fig:histogram} shows that the runaway fraction through this method depends very strongly on the chosen threshold.   

    
    \item {\bf Using the normalised peculiar velocities:} We calculate the uncertainty-normalised peculiar velocities as: 
    \begin{equation}
E=\sqrt{\left(\dfrac{V_b-\overline{V}_b}{3\overline{\sigma}_b}\right)^2+\left(\dfrac{V^{gal}_\ell-V_\ell^{model}}{3\overline{\sigma_{\ell}}}\right)^2}>1,
\label{eq:statistical}
    \end{equation}
\noindent where $\overline{\sigma}_b=\sqrt{\sigma_b^2+\Delta_b^2}$ and $\overline{\sigma}_\ell=\sqrt{\sigma_\ell^2+\Delta_\ell^2}$; $\sigma_b$ and $\sigma_\ell$ being the dispersion in their respective directions and $\Delta_b$ and $\Delta_\ell$ the uncertainty in the velocity of each star.


\end{itemize}

We consider the first method to be the most relevant, as it aligns more closely with the historical definition of runaway stars. The second method remains useful for identifying high-confidence runaways, as it excludes stars with large relative velocity errors. Implementing the second method also enables direct comparison with studies that adopt it as their primary selection criterion.

\label{sec:uncertainties}
The proper motion and distance uncertainties were propagated to obtain the uncertainties in the total velocities in the $b$ and $\ell$ directions, i.e. $V_b$ and $V^{gal}_\ell$. We run a Monte Carlo experiment consisting of 500 simulations drawing values for the velocities from an asymmetric normal distribution based on Eq. \ref{eq:asy_norm}, and drawing values for the parameters in the model from the last 100 steps of the MCMC run described in Section \ref{sec:fit}. Once the total velocities and the Galactic rotation model have been sampled, we can derive the peculiar velocities and apply both methods explained above.

For each star we calculated the probability of it being a runaway by dividing the number of times it was found to be a runaway star by either method divided by the total number of simulations.

For each simulation we calculate the fraction of stars that are flagged as runaways. The runaway fraction is calculated as the median of the value obtained for each simulation, and is reported with the $16^{th}$ and $84^{th}$ percentiles.

\subsection{Results}
\begin{center}
    \begin{figure}
        \centering
        \includegraphics[width=\linewidth]{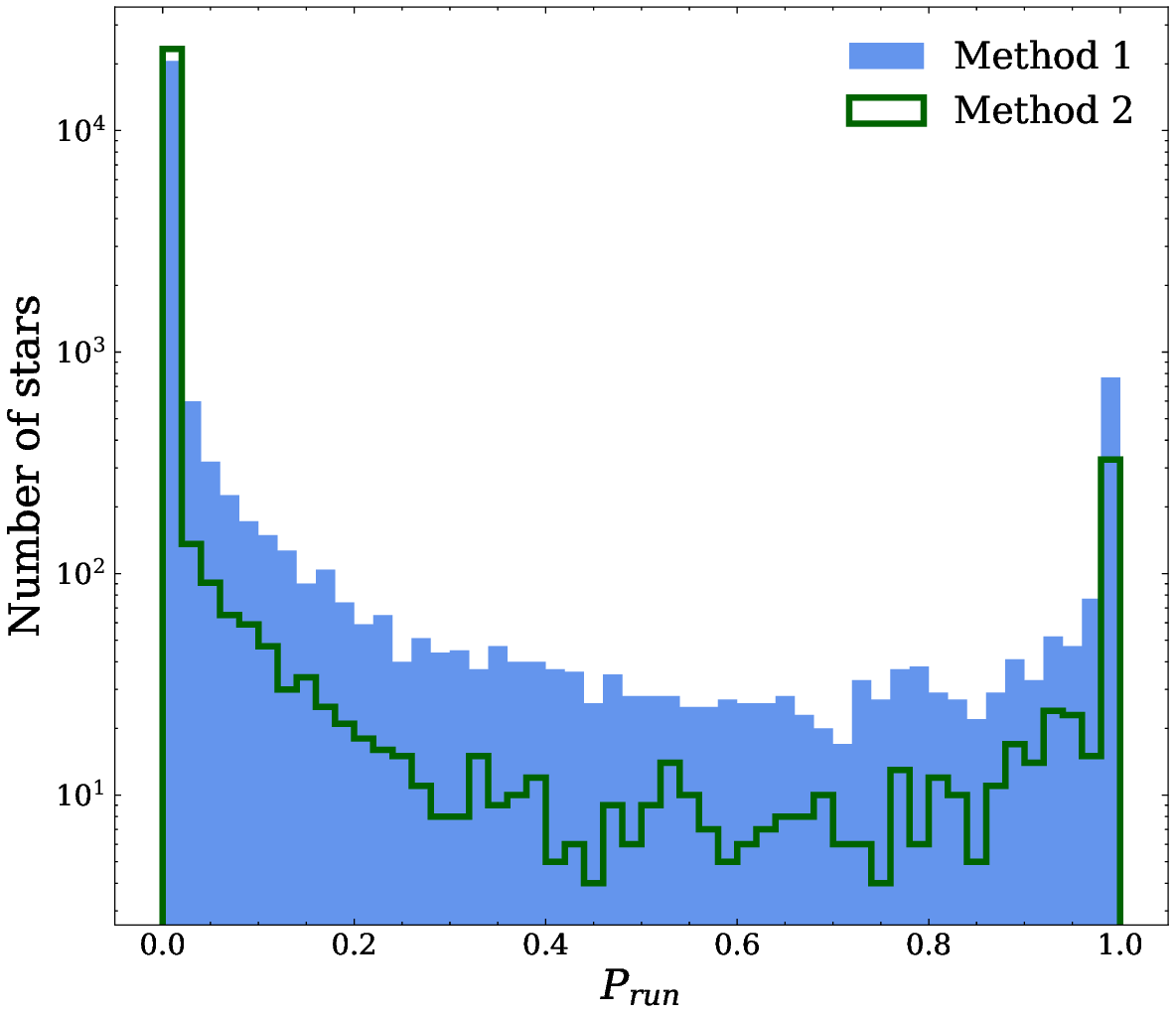}
        \caption{Distribution of runaway probabilities for methods 1 and 2.}
        \label{fig:prob_hist}
    \end{figure}
\end{center}

For method 1, we find a runaway fraction of $17.5^{+0.1}_{-2.5}\%$ for O-type stars and $7.0\pm0.1\%$ for B-type stars. For method 2 we find a runaway fraction of $10.0\pm2.5\%$ and $2.5^{+0.4}_{-0.3}\%$ for O- and B-type stars respectively.
Method 1 identifies more stars than method 2. This is particularly true for O-type stars, for which no star is selected as a runaway star by Method 2 that wasn't selected as a runaway star by Method 1. Figure \ref{fig:prob_hist} shows that the probabilities for either method are concentrated near 0 or 1. 

We choose $P_{run}=0.5$ as the threshold above which a star is classified as a runaway star. This threshold is used to classify individual stars, but not to calculate the runaway fraction, which is calculated using the Monte Carlo approach explained in Section \ref{sec:uncertainties}.

The fractions obtained by both methods also confirm that O-type stars are more likely to become runaways than B-type stars, as previous studies have hinted at, but which has never before been shown using a volume-complete sample.

\begin{table}
    \centering
    \caption{Top 10 O-type stars by runaway probability of both methods, $P^{(1,2)}_{run}$. Full list will be made available. $V$ represents the median peculiar velocity of each star with its associated uncertainty.}
    \label{tab:O_runaways}
    \renewcommand{\arraystretch}{1.3} 
    \begin{tabular}{llrrl}


\toprule
Name & HD &$P_{run}^{(1)}$  & $P_{run}^{(2)}$ & $V (\,km\,s^{-1})$ \\
\midrule
$\mu$ Col & 38666 & 1.000 & 0.918 & ${82.9}^{+29.6}_{-11.2}$ \\
AE Aur & 34078 & 1.000 & 1.000 & ${92.8}^{+1.7}_{-1.6}$ \\
$\zeta$ Pup & 66811 & 0.994 & 0.700 & ${37.6}^{+4.4}_{-6.6}$ \\
$\zeta$ Oph  & 149757 & 0.992 & 0.000 & ${26.3}^{+1.3}_{-1.3}$ \\
68 Cyg & 203064 & 0.986 & 0.810 & ${32.2}^{+1.5}_{-3.3}$ \\
$\lambda$ Cep & 210839 & 0.964 & 0.574 & ${43.2}^{+6.7}_{-10.1}$ \\
HD  93521 & 93521 & 0.700 & 0.002 & ${28.1}^{+4.7}_{-12.0}$ \\
$\xi$ Per & 24912 & 0.308 & 0.074 & ${18.6}^{+10.2}_{-4.2}$ \\
HD 159176 & 159176 & 0.024 & 0.000 & ${13.6}^{+4.6}_{-4.4}$\\
V986 Oph & HD 165174 & 0.006 & 0.000 & ${14.4}^{+3.6}_{-4.2}$ \\
\bottomrule

\end{tabular}
\end{table}
{
\subsection{Calibration of results with 3D kinematics}
Our calculations have been performed in 2D due to the high availability of 2D proper motions from Gaia and the low availability of radial velocities in the third dimension. In this section we calibrate our results by repeating our analysis on a subset of our sample for which 3D velocities exist.

To construct the sample of OB stars with 3D velocities, we cross-matched our catalogue of OB stars with spectroscopic data from APOGEE DR17 \citep[][]{APOGEE16, APOGEE22} and LAMOST DR6 \citep[][]{LAMOST, LAMOST2022}. Our cross-matched sample comprises 3,884 stars with 3D velocities, with a median relative uncertainty on the radial velocities of $5.4\%$

To compute the peculiar radial velocities, we remove the contributions from the Sun’s motion ($ V_{R,\odot}$, $V_{\phi,\odot}$, $V_{Z,\odot}$), the mean motion in the Galactic plane and the mean vertical ($Z$) motion, after projecting each component onto the solar radial direction. The mean motion on the Galactic plane can be obtained using the best fit parameters we obtained fitting the proper motions (see Table \ref{tab:best_fit}). We find that the vertical component projected onto the Solar radial direction ($V_{Z}\sin{b}$) does not depend significantly on $b$ or $\ell$, and is therefore calculated as the median radial velocity once the other components have been subtracted, obtaining a value of  $\overline{V_{Z}\sin{b}}=2.80\, km\, s^{-1}$.

Taking all of this into account, the peculiar radial velocity can the be calculated as:

\begin{equation}
\begin{aligned}
V_r^{\mathrm{pec}} = V_r 
&- \left( V_{R,\odot}\cos\ell - V_{\phi,\odot}\sin\ell \right)\cos b 
- V_{Z,\odot}\sin b \\
&- \left( V_\phi \sin(\phi+\ell) - V_R\cos(\phi+\ell) \right)\cos b \\
&- \overline{V_{Z}\sin b}.
\end{aligned}
\end{equation}

 Figure \ref{fig:pec_rv} shows a comparison between the measured radial velocities and the velocities predicted by the Galactic model. There is a clear general trend that agrees well, albeit with a large dispersion around the 1-1 trend line.

\begin{center}
    \begin{figure}
        \centering
        \includegraphics[width=\linewidth]{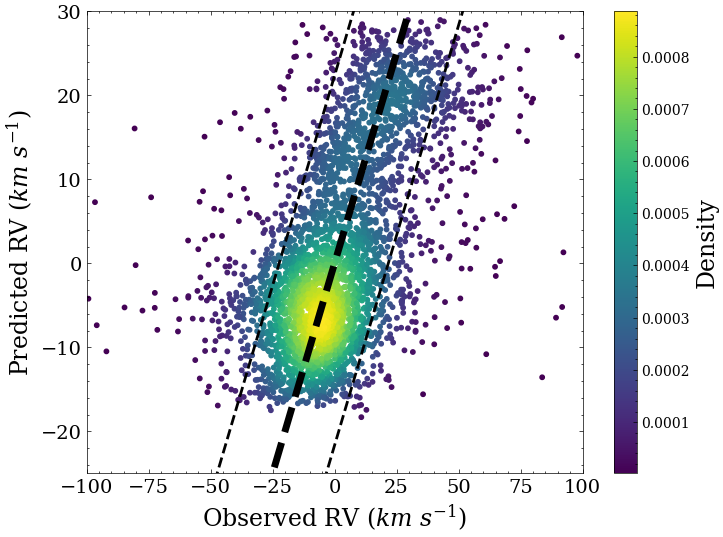}
        \caption{ Comparison between the observed radial velocities and the radial velocities predicted using the solar motion, the Galactic rotation model and the mean vertical motion. The thick black line corresponds to the 1-1 line, while the thinner black lines are parallel to the 1-1 line set at the 16th and 84th percentiles of the peculiar radial velocities ($-15.8\,km\,s^{-1}$ and $15.3\,km\,s^{-1}$ respectively) calculated as the observed minus de predicted radial velocities.}
        \label{fig:pec_rv}
    \end{figure}
\end{center}

We fit the 3D velocity distribution to the Maxwell-Boltzmann distribution given by Eq. \ref{eq:Boltzmann_spatial} using stars with total velocities up to $20\,km\,s^{-1}$ to avoid fitting the tail, as it is a combination of the Maxwell-Boltzmann distribution and the excess velocity stars. We obtain best-fit parameters of $a_{3D}= (7.1\pm 0.2)\times 10^{-4} \,(km\,s^{-1})^{-3}$ and $b_{3D} = 13.9\pm0.2 \, km\,s^{-1}$. The value of $b$ is slightly higher than the one we obtained for the 2D distribution ($b=10.99\pm 0.06$), which likely stems from the less accurate determination of the peculiar velocities. The histogram of 3D velocities and the Maxwell-Boltzmann fit are shown in figure \ref{fig:3dhist}.

Applying the same criterion used for the 2D velocities, we obtain a runaway velocity threshold of $28\,km\,s^{-1}$. For a perfectly isotropic sample we would expect the 3D velocity threshold to be $\sqrt{\dfrac{3}{2}}$ times the 2D one. In this case it would be $23\,km\,s^{-1}\times\sqrt{\dfrac{3}{2}} \approx 28.2\,km\,s^{-1}$. These two threshold values are in good agreement, suggesting that our 2D velocity threshold and selection of runaway stars in 2D provides a reasonable approximation to a full 3D treatment. Note that there is an uncertainty on this value associated with the bin size used on each histogram ($\pm0.5\,km\,s^{-1}$ in the 2D one and $\pm 1\,km\,s^{-1}$ for the 3D value). Using stars with velocities beyond $20\,km\,s^{-1}$ in the fit widens the distribution and can therefore increase the runaway selection threshold.

\begin{center}
    \begin{figure}
        \centering
        \includegraphics[width=\linewidth]{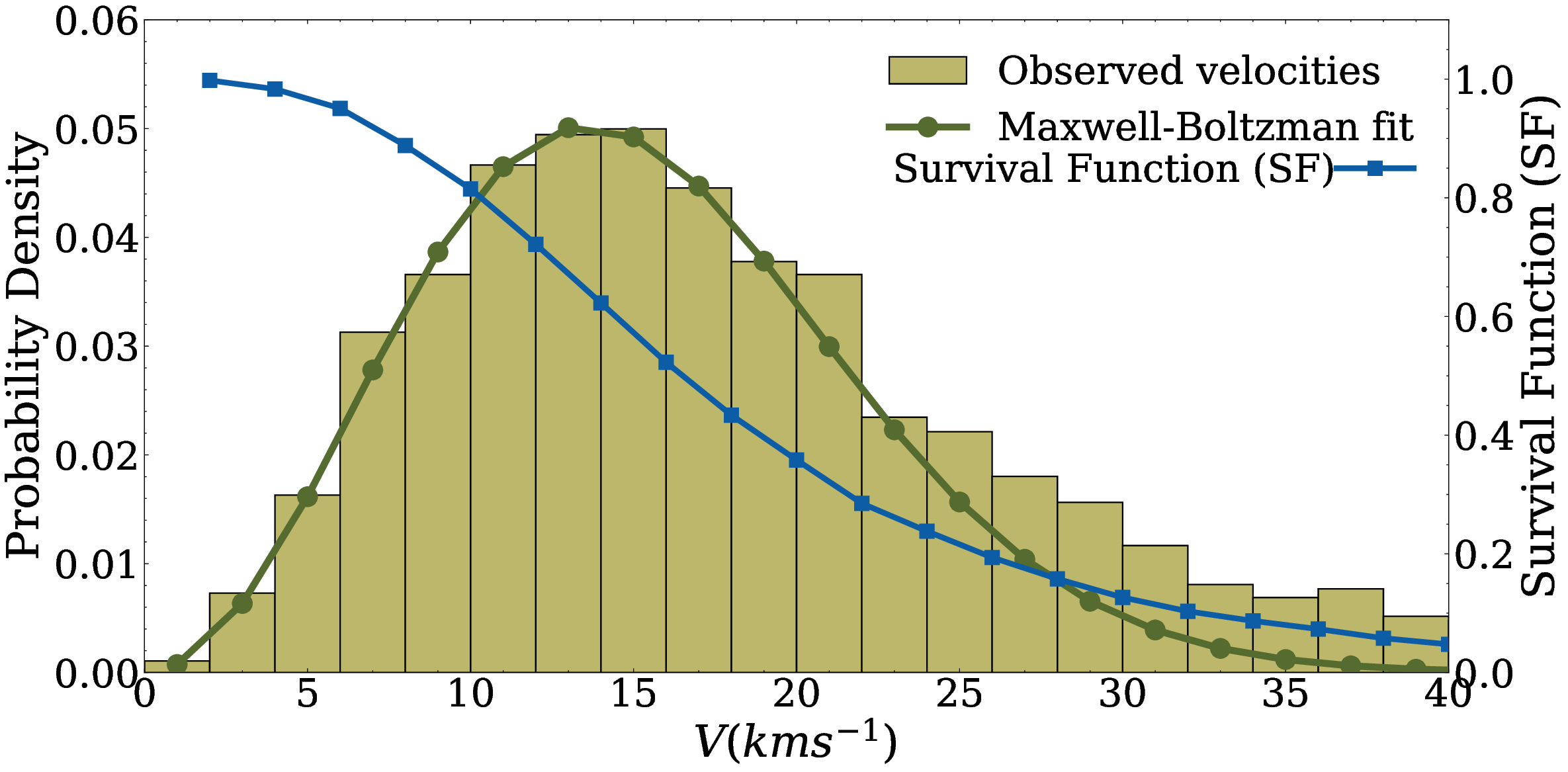}
        \caption{ Histogram showing the distribution of 3D residual velocities defined as $V=\sqrt{(V_\ell^{gal}-V_\ell^{model})^2+(V_b-\overline{V}_b)^2+(V_r^{pec})^2}$. Also shown are the Maxwell-Boltzmann fit to this distribution, represented by the dark green line marked with circles, and the Survival Function, represented by the blue line marked with circles.}
        \label{fig:3dhist}
    \end{figure}
\end{center}}
\section{Discussion}

\label{sec:literature_comparison}

In this section we discuss our results, first by considering the runaway status of some of the most common runaway star candidates in the literature, and then by comparing our overall statistics with other observational and theoretical studies in the literature.

\subsection{Comments on individual O-type stars}
The low number of O-type stars within 1 kpc of the Sun coupled with their high luminosity has meant that many runaway O-type stars have been identified in the past based on either their location, proper motion or radial velocity. We discuss the individual runaway stars we have identified here (shown in Table \ref{tab:O_runaways}):
\begin{itemize}
    \item $\mu$ Col, AE Aur, $\zeta$ Oph, 68 Cyg and $\lambda$ Cep were identified as runways by \cite{Blaauw1961} based on their radial velocities. We confirm the runaway status of these stars here based on their proper motions, particularly using Method 1, but note that Zeta Oph is not classified as a runaway star using Method 2. 
    \item $\zeta$ Pup was, to our knowledge, first proposed as a runaway star by \cite{Upton1973}. We confirm it to be a runaway star here using both methods applied to the proper motions.
    \item HD 93521 was first proposed as a runaway star by \cite{Hack1977} based on its radial velocity. Based on its proper motion we find it to be a runaway star using Method 1, but not using Method 2.
\end{itemize}

Among the stars that we find unlikely to be runaway stars, only {\it X} Per and $\xi$ Per have been consistently considered a runaway star in the literature. $\xi$ Per is below the threshold probability for methods 1 and 2, but still has a moderate probability of being a runaway star from Method 1. Many authors \citep[e.g.][]{Blaauw1961} have identified $\xi$ Per as runaways based on its radial velocity, but clearly its proper motions are not high enough for it to stand out as a runaway star based on the proper motion alone. \citet{Huthoff2002} used the Simbad database to calculate the peculiar radial velocity of X Per, obtaining $-32.0\,km\,s^{-1}$, making it a runaway star. The peculiar 2D velocity they find based on its Hipparcos proper motion, $13.4 \,km\,s^{-1}$, is similar to the one we find ($11.2^{+1.9}_{-1.8}\,km\,s^{-1}$), which is not high enough to classify this star as a runaway star. 
In these two cases, 2D velocities are not sufficient to determine their runaway status. It is to be expected that a number of runaway stars will not have been identified as such when analysing only their proper motions, but the overall statistics of the sample should be similar if we assume spatial isotropy (which was already assumed to reach Eq. \ref{eq:Boltzmann_parallel}). { Conversely, some stars classified here as runaways would not meet a 3D velocity threshold derived using the same approach, since the higher threshold would exclude objects with low peculiar radial velocities.}

\cite{Noriega-Crespo1997} studied excess $60\mu m$ emission around OB stars to find stars with possible bowshock nebulae around them. Their observations of 10 Lac suggest that it could have a bowshock nebula. HD 207538, HD 24431, HD 199579 also show excess $60\mu m$ emission, but the structure cannot be sufficiently resolved to identify a bowshock nebula. However, the proper motions of these stars are not high enough to suggest that any of them is a runaway. Note that bowshock nebulae can be produced by non-runaway stars \citep[e.g.,][]{Kobulnicky2022}.

\subsection{Spatial distribution}

\begin{figure*}[t]   
\centering

\begin{minipage}{0.48\textwidth}
    \centering
    \includegraphics[width=\linewidth]{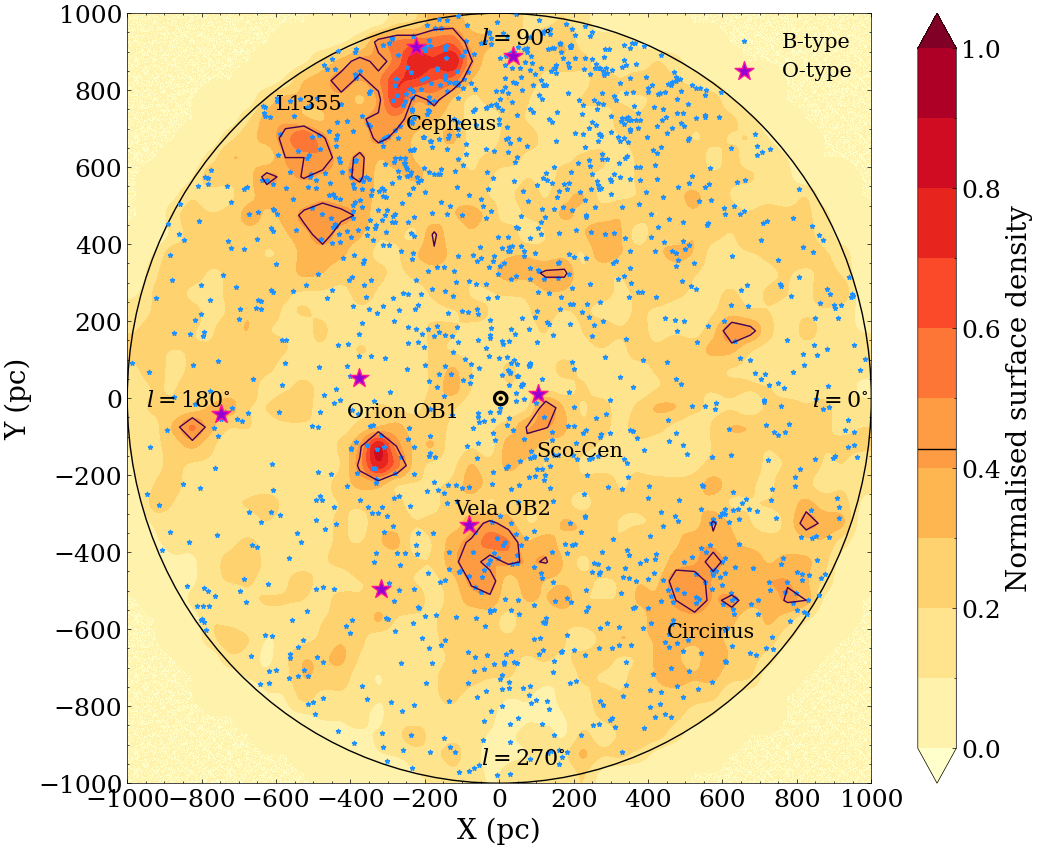}
\end{minipage}
\hfill
\begin{minipage}{0.48\textwidth}
    \centering
    \includegraphics[width=\linewidth]{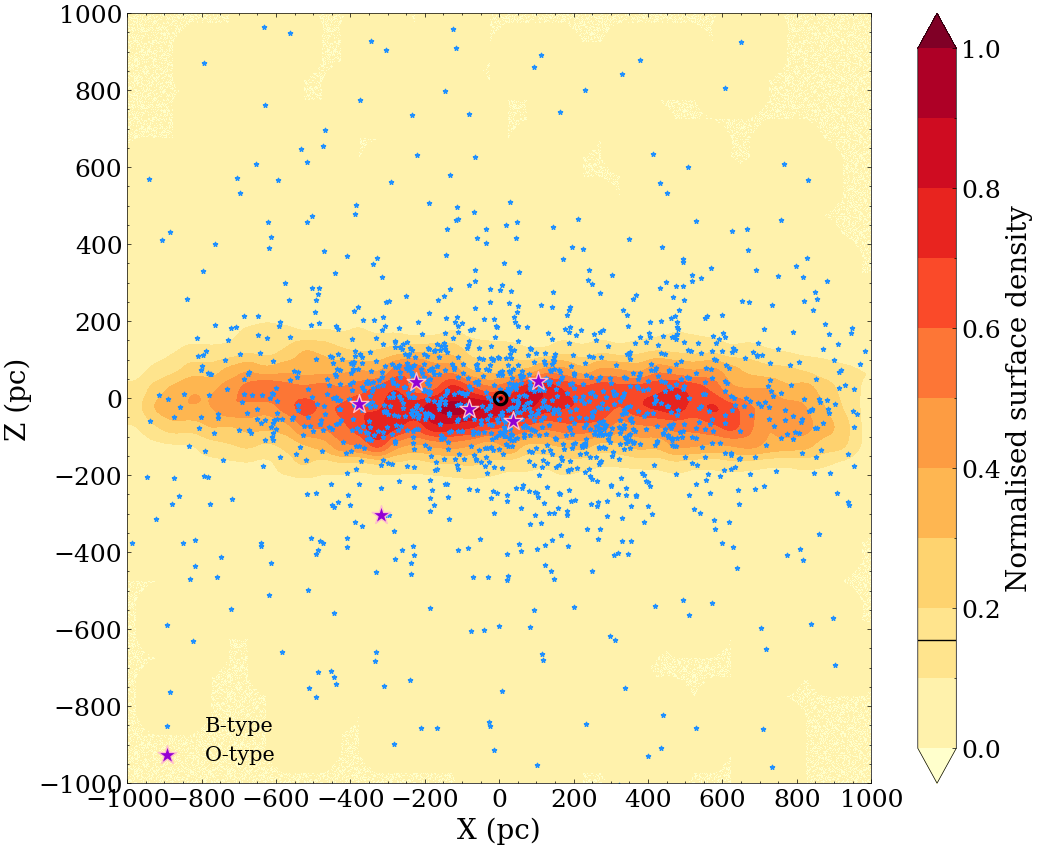}
\end{minipage}

\vspace{0.5cm}

\begin{minipage}{0.48\textwidth}
    \centering
    \includegraphics[width=\linewidth]{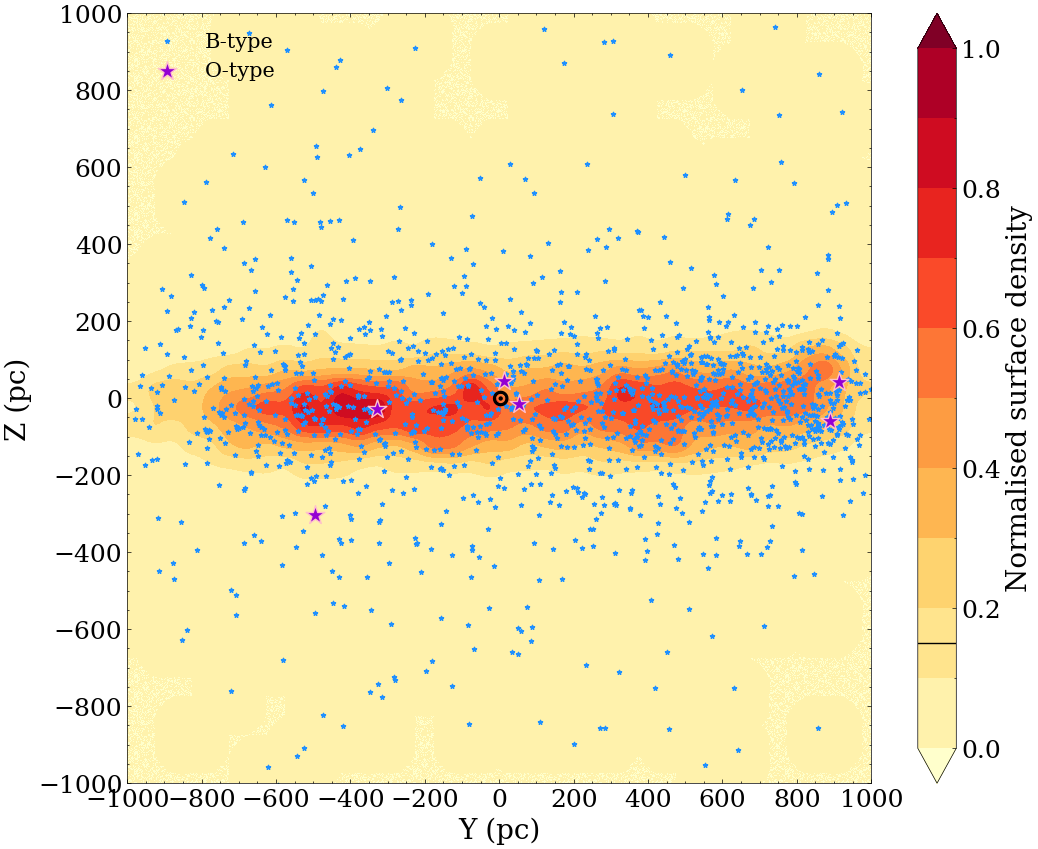}
\end{minipage}
\hfill
\begin{minipage}{0.48\textwidth}
    \centering
    \includegraphics[width=\linewidth]{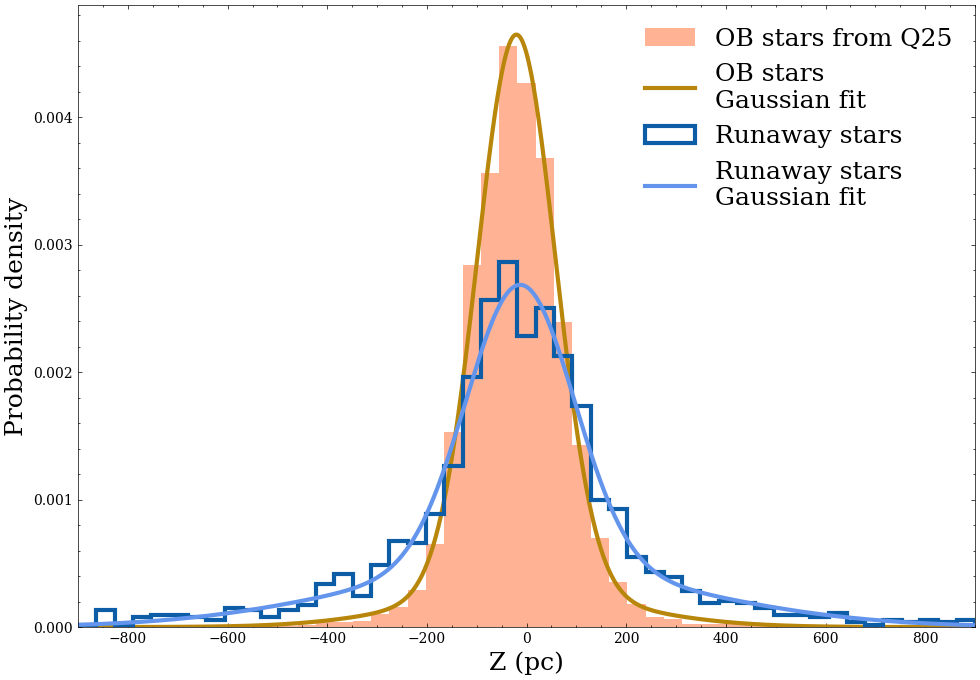}
\end{minipage}
\label{fig:runawaysXY}
\caption{Panels (1)-(3): Distribution of the runaway stars selected with Method 1 in (X,Y,Z), plotted over the spatial density of OB stars from \citetalias{Quintana25}. Panel (4) Gaussian Fit of the Z distribution of our runaway stars and the OB stars from \citetalias{Quintana25}.}
\end{figure*}

{ 
Figure \ref{fig:runawaysXY} shows the distribution of runaway stars in the X-Y plane of the Galaxy, along with some of the most prominent OB associations. The regions around the Cepheus complex show the largest concentration of B-type runaway stars. Most O-type stars are close to a large OB association. This is expected as they are younger and less likely to be produced by less massive clusters or associations.

A census of OB associations within 1kpc of the Sun has recently been published \citep{Quintana2026}, which will allow us to perform a more detailed analysis on the origin of our runaway stars in a follow-up work.

Panels (2) and (3) of Figure \ref{fig:runawaysXY} show the X-Z and Y-Z distribution of runaway stars. The distribution of B-type runaway stars is more diffuse than the general distribution of OB stars. O-type runaway stars, on the other hand, are located closer to the Galactic plane. This is again likely due to their lower age.

We fitted the Z-distribution to a combination of Gaussians of the shape:

\begin{equation}
    \mathcal{F}(Z)=\sum_{i=1,2}W_i\mathcal{G}(Z;\bar{Z},\sigma_i)\ ,
    \label{Eq:Zdis}
\end{equation}
\noindent were $\mathcal{G}(Z;\bar{Z},\sigma_i)$ is a Gaussian with mean $\bar{Z}$ and standard deviation $\sigma_i$, and $W_1+W_2=1$.
\begin{table}
    \centering
        \caption{Best fit parameters for Eq. \ref{Eq:Zdis}.}

    \begin{tblr}{hline{1,3,5} = {-}{}, colspec = {lcccc}
    }
        Population&$W_1$ & $\bar{Z}$& $\sigma_1$ & $\sigma_2$  \\
        & & pc& pc & pc  \\
        \citetalias{Quintana25} OB stars & 0.95 & -19.79 & 84.6 &299.5\\ Runaway stars & 0.61 & -12.83& 108.2& 347.1
         
    \end{tblr}
    \label{tab:Zdis}
\end{table}

Table \ref{tab:Zdis} shows the best fit parameters for the runaway OB stars and the whole population of OB stars. For runaway stars, the dispersion of both the wide and thin components is slightly larger than the whole OB star sample. For runaway OB stars, the thin component is much less prominent, with a weight of $0.61$ compared to $0.95$ in the general population. This means that stars outside the Galactic plane are more likely to be runaway stars.}

\subsection{Comparison with the literature}

\begin{table*}
\centering
\caption{Comparison of the runaway fraction obtained by various authors and those found on this work. The runaway fraction is calculated as $\dfrac{N_{\rm runaways}}{N_{\rm stars}}$ except for the values in this work, which are calculated following Section \ref{sec:uncertainties}. The Poisson noise is calculated using Eq. \ref{Poisson}. \label{comparisonrunaways}}
\begin{tblr}{
  hline{1-2,7,19} = {-}{},
}
Catalogue                      & Type         & $N_{\rm runaways}$ & $N_{\rm stars}$ & Fraction             & Poisson Noise \\
{\bf This work:} \\
Method 1       &
                               O            &   7               &       40    & $17.5^{+0.1}_{-2.5}\%$          &             $6.6\%$  \\
                               & B            &   1502               &      24,488     & $7.0\pm 0.1\%$ & $0.2\%$              \\
                               Method 2       
                                 & O           & 5                &    40       &   $10.0\pm2.5\%$        &   $5\%$   \\
                               & B            &   861               &      24,488     & $2.5^{+0.3}_{-0.4}\%$ &  $0.2\%$                   \\
\cite{Guo2024} 2D             & OB           & 480              & 4432      & 10.8\%               & 0.5\%         \\
\cite{Guo2024} 3D             & OB           & 229              & 4432      & 5.2\%                & 0.3\%         \\
\cite{Carretero-Castrillo2023} & O2-O7        & 50               & 199       & 25.1\%               & 3.6\%         \\
                               & O7-O9        & 46               & 194       & 23.7\%               & 3.5\%         \\
                               & B0e-B3e      & 36               & 585       & 6.2\%                & 1\%           \\
                               & B4e-B9e      & 23               & 482       & 4.8\%                & 1\%           \\
\citet{Drew2021}             & O ($20\,km\,s^{-1}$)            & 102               & 733       & 13.9\%                & 1.4\%         \\
                               & O ($25\,km\,s^{-1}$)         & 75               & 733       & 10.2\%                & 1.2\% \\ & B ($20\,km\,s^{-1}$)           & 218               & 2512       & 8.7\%                & 0.6\%         \\
                               & B ($25\,km\,s^{-1}$)           & 136               & 2512       & 5.4\%                & 0.5\% \\
\cite{MaizApellaniz2018}            & O            & 48               & 871       & 5.5\%                & 0.8\%         \\
                               & B            & 27               & 959       & 2.8\%                & 0.5\%

\end{tblr}
\label{tab:fractions}
\end{table*}

In this section, we compare the runaway fraction we obtained to recent measurements in the literature with similar selection criteria.

Since most runaway fractions in the literature do not include uncertainties, we calculated the Poisson noise as a basic measure of the uncertainties associated with each published value:
\begin{equation}
    \Delta X= \dfrac{\sqrt{N_{runaways}}}{N_{stars}}
    \label{Poisson}
\end{equation}
Note that the Poisson noise just measures the uncertainties associated with randomness in a finite sample size, and therefore is a lower limit to the real uncertainty, which can be inflated by many other causes.

Table \ref{comparisonrunaways} lists measures of the runaway OB star fractions in the literature from various studies. A wide variety of different values have been measured, from $2.8$ to $25.1\%$. Differences in these values arise from the different samples used (which cover different ranges of stellar spectral types) and different methods for selecting runaway stars.

\cite{MaizApellaniz2018} identified runaway stars from three different samples of OB stars: O stars from the Galactic O-Star Spectroscopic Survey (GOSSS), Stars labelled as O-type on SIMBAD and stars labelled as BA supergiants on SIMBAD. Their spectroscopic sample is therefore neither volume nor magnitude complete and highly heterogeneous. They identified runaway stars by comparing the proper motions (from \textit{Gaia} DR1 and HIPPARCOS) of each star with that of the whole sample, but contrary to this paper (and also \cite{Carretero-Castrillo2023}, discussed below), they do not iterate, which results in lower runaway star fractions. Out of the 76 runaway stars identified by \cite{MaizApellaniz2018}, 13  are within 1 kpc in \citetalias{Quintana25}. We find that 9 of them are runaway stars according to our criteria, but the other 4 (v {\it cen}, $\kappa$ {\it Cas}, $\theta$ {\it Ara}, $\phi$ {\it Vel}) have peculiar velocities that are too low to be considered runaway stars. 

\cite{Carretero-Castrillo2023} used 2D kinematics (\textit{Gaia} DR3 proper motions) and selected runaway stars by iteratively applying Eq. \ref{eq:statistical} to a sample of OB stars from GOSC and the Be Star Spectra \citep[BeSS, ][]{Bess2011} database. Again, this means that their underlying sample is neither volume or magnitude complete. Comparing their results to our Method 2 results, their runaway fractions are higher for both O-type stars ($23.7-25.1\%$ versus $10.0\%$) and B-type stars ($4.8-6.3\%$ versus $2.5\%$; see Table \ref{tab:fractions}).  The disagreement in the O-type star runaway fractions between their work and ours might be caused by a selection bias. Runaway O-type stars are more likely to be spectroscopically observed than their non-runaway counterparts, because they spend a larger portion of their lifetimes in a non-embedded environment. This means that a spatially incomplete sample is likely to be biased towards runaway stars, as this comparison suggests. Our use of a spatially complete sample of O-type stars overcomes this bias.

Although we do not expect this bias to be significant for the B-type stars, the fraction we obtain using method 2 is lower than the values obtained by \citet{Carretero-Castrillo2023}, even accounting for uncertainties. Our value ($2.5\%$) is closer to the value that \citet{Carretero-Castrillo2023} obtained for mid-to-late B-type stars ($4.8 \pm 1.0\%$). Given that our sample of B-type stars is dominated by late B-type stars, this is not surprising.




Only one O-type runaway star and 27 B-type runaway stars found by \cite{Carretero-Castrillo2023} lie within $1kpc$ of the Sun. Out of these, 17 B-type stars were included in \citetalias{Quintana25}. The mean runaway probability for these objects is $71.6\%$, which indicates a generally good agreement, while three of them were found unlikely to be runaway stars ({\it HD 7636}, {\it HD 90436} and {\it HD 140114}).

\cite{Guo2024} identified runaway OB stars using both 2D (\textit{Gaia} DR3 proper motions) and 3D velocities (\textit{Gaia} proper motions plus radial velocities from the Large sky Area Multi-Object fiber Spectroscopic Telescope (LAMOST) DR8), due to the fact that their radial velocities were mostly based on single observations and therefore susceptible to error from unresolved binaries. Their sample of 4432 OB stars is based on the Large sky Area Multi-Object fiber Spectroscopic Telescope (LAMOST) Data Release 8. Again, it is neither volume nor magnitude complete.

\cite{Guo2024} identified runaway OB stars using a set velocity threshold, set to be $1\%$ of the peak of their fitted Maxwell-Boltzmann distribution ($23.87\,km\,s^{-1}$ and $42\,km\,s^{-1}$ for the 2D and 3D methods respectively). Our method 1 is the most similar to their 2D method and uses a very similar velocity threshold.

\cite{Guo2024} estimate that around $3\%$ of the stars of their sample have an effective temperature over $30,000$ K to be compared with the $0.2\%$ O-type star fraction for our sample. Since their sample is dominated by B-type stars we can still compare to our runaway star fraction for B-type stars. Since our method is 2D-only, we compare it to their 2D value. Our runaway star fraction of B-type stars (Method 1, $7.0\%$) is notably smaller than the value they obtain ($10.8\%$). Even when we take into account the sample composition, the weighted mean of our O and B-type star runaway fraction is only $7.3\%$, still bellow their 2D value.

{The runaway fractions obtained by \citet{Drew2021} from a sample of 4199 OB stars in the Carina Arm can be compared to the ones found in this paper. For O-type stars, they found a runaway fraction of $10.2\pm1.2\%$ using a peculiar velocity threshold of $25\, km \, s^{-1}$  and $13.9\pm1.4\%$ with a $20 \, km \, s^{-1}$ threshold, both of which are lower than the value we obtained, $17.5^{+0.1}_{-2.5}\pm6.6\%$, using a threshold of $23\, km \, s^{-1}$. These values are still compatible when we take into account the large Poisson uncertainty driven by the small sample size of O-type stars within $1\, kpc$ of the Sun. The B-type star runaway fractions found by \citet{Drew2021} are $8.7\pm0.6\%$ and $5.4\pm0.5\%$ for thresholds of $20$ and $25\, km \, s^{-1}$ respectively. The fraction we obtained, $7.0\pm 0.1\%$ is in-between these two values, which shows good agreement with the two values, even considering that their sample contains only B0-B3- type stars, from which we would expect a slightly higher runaway fraction.}

\subsection{Comparison with simulations}

For O-type stars, BSS simulations \citep[]{Portegies2000, Renzo2019} suggest that the runaway fraction is $2\%$ at most, which is well below our measured value. This suggests that the majority of runaway O-type stars are likely to have been produced by the DES mechanism.

N-body simulations can predict the runaway fraction from the DES mechanism. Model 009 by \cite{Perets2012} (the most realistic  of their models) predicts that a runaway fraction of $\sim7\%$ for early B-type stars when the velocity threshold is set to $20\,km\,s^{-1}$. This value rapidly drops to less than $1\%$ for stars below $10M_\odot$. The same model predicts a runaway fraction of $10$ to $20\%$ for O-type stars. \citet{Oh2016} ran a range of models with different binary properties (period distribution, eccentricity) and cluster properties (half-mass radius, initial mass segregation) that predict B-type runaway fractions of $0-4\%$, most of them predicting about $1\%$, and O-type runaway fractions of $3-10\%$ (peculiar velocities over $30\,km\,s^{-1}$). The low runaway fraction of early B-type stars found by both \citet{Perets2012} and \citet{Oh2016} could indicate that many of the runaway B-type stars we find in this paper were produced via BSS, while the O-type star runaways are mainly produced via DES.

{

\subsection{Effects of the local structure of the Milky Way} 
In our model, we have assumed that the motion of stars in the Milky Way is axisymetrical. However, the local structure of the Milky Way may affect their kinematics.

\begin{itemize}
    \item {\bf Local Bubble:} The Local Bubble is a region around the Sun where the density of stars and gas is lower compared to other regions in the Solar Neighbourhood \citep[e.g.][]{Cox1987}. However, the Local Bubble is not present in the spatial distribution of OB stars when a volume-complete sample is used \citep[see][]{Quintana25}, so we do not expect kinematic variations due to this effect.
    \item {\bf Asymmetric drift:} The distribution of Galactic rotation velocities is not symmetric, as there is a long tail of stars that rotate slower \citep[see Chapter 4.8.2 of ][]{Binney2008GalacticDynamics}. Young stars, and in particular massive stars, have a small asymmetric drift. Applying the fit performed by \citet[]{Almannaei2024} to our sample, we find that the asymmetric drift is of the order of $1\,km\,s^{-1}$ or smaller for all of our objects.
    \item {\bf Dependence of peculiar velocities with $\phi$:} Given the short range of $\phi$ values covered by our sample (approximately -0.12 to 0.12 rad) we do not expect a significant dependence beyond the effects of the projection (e.g. Equation \ref{projection}).

    \item {\bf Other kinematic structures:} The lack of radial velocities for all stars in our samples means that we do not have complete information on the azimuthal velocity towards $\ell=90\degree,270\degree$ and incomplete information on the Galactic radial velocity of stars towards $\ell=0\degree,180\degree$, which might hide other kinematic structures and affect the fit parameters in Table \ref{tab:best_fit}. These effects, however, do not affect the calculated peculiar velocities in the $\ell$ direction as they disappear with the projection (i.e. Eq. \ref{projection}).
\end{itemize}
}

\section{Conclusions}
\label{conclusions}

We have used a volume-complete sample of OB stars within 1 kpc of the Sun to identify and characterise runaway OB stars. 

We fitted a Galactic rotation model to the sample of OB stars and subtracted these from the observed motions to calculate peculiar velocities. We find that the peculiar velocities are well represented by a Maxwell-Boltzmann distribution with an excess probability density at high velocities that we believe are dominated by OB stars ejected from their birth cluster.
    
We selected runaway stars by applying two methods: firstly by applying a fixed 2D peculiar velocity threshold; and secondly by selecting runaways as those whose velocities are significantly higher than the sample dispersion, accounting for measurement error. { We calibrated these methods by performing a similar analysis in 3D using a subset of stars with available RVs. }

Using these two methods and a Monte Carlo simulation, we derived the probability that each of the 40 O-type stars and 24,488 B-type stars in our sample is a runaway star. We also calculated the fraction of stars that are runaway stars for O- and B-type stars, $17.5^{+0.1}_{-2.5} \%$ and $7.0\pm0.2\%$ respectively, using the first of our two methods, which we argue is the more consistent approach compared to past works. The second method gives lower numbers of runaway stars. Although these values are similar to the ones obtained by other previous studies, there are some differences, which can be attributed mainly to the selection method and the underlying sample used, with most samples being either magnitude-limited or based on a spectroscopic sample, both of which can introduce biases. Our volume-complete sample overcomes this issue.

In a follow-up work we will try to trace the runaway stars identified in this paper to find their possible birth clusters or associations, and therefore calculate precise ejection velocities.

\section*{Acknowledgements}

This work was supported by the STFC and the Faculty of Natural Sciences at Keele University.

This work made use of the python packages Astropy \citep{astropy:2013, astropy:2018, astropy:2022}, Astroquery \citep{Astroquery2019}, Numpy \citep[][]{Numpy}.

We thank the anonymous reviewer for their helpful suggestions.

\section*{Data Availability}

The sample of O-type stars is available in Table \ref{tab:True_True}.
The full list of selected runaway stars and their probability will also be available.



\bibliographystyle{mnras}
\bibliography{references} 



\appendix

\section{O stars within 1kpc}

In Table \ref{tab:True_True} we present all O-type stars coming from three lists: (1) Stars both in GOSC and \citetalias{Quintana25}, (2) Stars in GOSC with astrometric distance smaller than 1 kpc and (3) stars with hogh effective temperature in \citetalias{Quintana25} not present in GOSC. In Table \ref{tab:rejected} we present all of the stars from lists (2) and (3) that were not included in the final sample.

\begin{table*}
\caption{Candidate O-type stars from GOSC (first part) and \citetalias{Quintana25} (second part) rejected from our final list of O-type stars within 1 kpc of the Sun (\ref{tab:True_True}). The source of the distance is \citetalias{Quintana25} unless stated otherwise: \dag \citet{Bailer-Jones2021}; \ddag parallax-inversion distance. Reasons for exclusion: a. Astro-photometric distance larger than 1 kpc, even if parallax distance suggests otherwise.  b. RUWE >8. c. Parallax error too large ($\varpi/\Delta \varpi<2$). d. Literature spectral type is not O. For GOSC spectral type references see description of table \ref{tab:True_True}.}
\label{tab:rejected}
\renewcommand{\arraystretch}{1.15} 
\begin{tabular}{llrrllll}
\toprule
HD number & Star Name & $D_{XY}$ (pc) & $T_{\rm eff} (10^3K)$ & Spectral Type & Spectral Type ref & Reason for exclusion \\
\midrule
17 505  & HD 17 505 A             & $877 \pm 631$ \ddag & & O6.5 IV     & M19a  &c \\
17 520  & HD 17 520 A             & $259^{+37}_{-36}$ \dag && O8 V      & M16a  &b \\
55 879  & HD 55 879               & $1038 ^{+270} _{-173}$ && O9.7 III    & S11a  & a \\
149 038 & mu Nor                  & $1035 ^{+299}_{-106}$ &&O9.7 Iab    & S14   & a \\
162 978 & 63 Oph                  & $982^{+182}_{-92}$ \dag && O8 II     & S14  & c \\
170 097 & HD 170 097 A            & $632^{+238}_{-120}$ \dag && O9.5 V      & M19a  & b \\
193 322 & HD 193 322 AaAb         & $1070 ^{+19}_{-13}$ && O9 IV     & S11a  & a \\
194 649 & HD 194 649 AB           &$837_{-42}^{+58}$\dag & &O6V&M19A & b \\
207 198 & HD 207 198              & $1025 ^{+126}_{-101}$ & &O8.5 II     & M16a   & a \\
209 975 & 19 Cep                  & $1182 ^{+136} _{-120}$ & &O9 Ib     & S11a  & a \\
\midrule
26 961 & * b Per & $998^{+957}_{-209}$ & $38^{+8}_{-8}$ & A1III & 2009ApJS..180..117A & d \\
175 687 & * ksi01 Sgr & $728^{+191}_{-124}$ & $31^{+7}_{-9}$ & B9/A0Ib   & 1988MSS...C04....0H & d \\
212 593 & *   4 Lac & $869^{+179}_{-133}$ & $37^{+10}_{-11}$ & B9Iab-Ib & 2024A\&A...690A.176N & d \\
205 139 & HD 205139 & $953^{+136}_{-123}$ & $32^{+10}_{-7}$ & B1Ib & 1968ApJS...17..371L & d \\
213 087 & *  26 Cep & $964^{+58}_{-17}$ & $42^{+5}_{-8}$ & B0.5Ib & 2024A\&A...690A.176N & d \\
206 165 & *   9 Cep & $965^{+285}_{-128}$ & $39^{+8}_{-9}$ & B2Ib & 2024A\&A...690A.176N & d \\
216 014 & V* AH Cep & $791^{+71}_{-32}$ & $31^{+2}_{-3}$ & B0.2V+B2V & 1997ApJ...490..328B & d \\
12 301 & *  53 Cas & $997^{+111}_{-102}$ & $42^{+7}_{-11}$ & B8Ib & 2024A\&A...690A.176N & d \\
36 881 & HD  36881 & $1000^{+460}_{-333}$ & $34^{+11}_{-11}$ & B9IIIp & 1972AJ.....77..750C & d \\
45 995 & HD  45995 & $783^{+34}_{-20}$ & $31^{+3}_{-6}$ & B1.5Vne & 1983ApJ...272..182A & d \\
37 490 & * ome Ori & $556^{+187}_{-74}$ & $34^{+4}_{-5}$ & B3Ve & 2006MNRAS.371..252L & d \\
37 660 & HD  37660 & $721^{+89}_{-73}$ & $32^{+7}_{-24}$ & A2II & 1999MSS...C05....0H & d \\
23 180 & * omi Per & $335^{+99}_{-20}$ & $33^{+6}_{-7}$ & B1III & 1973ARA\&A..11...29M & d \\
23 848 & * n Per & $972^{+726}_{-436}$ & $37^{+9}_{-6}$ & A3V & 1969AJ.....74..375C & d \\
23 338 & * q Tau & $920^{+989}_{-442}$ & $37^{+9}_{-9}$ & B6IV & 2024A\&A...690A.176N & d \\
33 034 & HD  33034 & $678^{+82}_{-63}$ & $31^{+5}_{-7}$ & A1II & 1965LS....C05....0H & d \\
35 600 & HD  35600 & $958^{+174}_{-133}$ & $34^{+10}_{-17}$ & B9Ib & 2024A\&A...690A.176N & d \\
29 479 & * sig01 Tau & $766^{+573}_{-683}$ & $33^{+10}_{-25}$ & A4/A5/A7 & 1995ApJS...99..135A & d \\
26 912 & * mu. Tau & $857^{+728}_{-409}$ & $37^{+9}_{-9}$ & B3IV & 2020A\&A...639A..81B & d \\
74 272 & * n Vel & $964^{+96}_{-68}$ & $38^{+10}_{-18}$ & A5II & 1978MSS...C02....0H & d \\
64 760 & * J Pup & $842^{+249}_{-208}$ & $35^{+10}_{-8}$ & B0.5Ib & 2015A\&A...580A..59R & d \\
74 195 & * omi Vel & $733^{+298}_{-340}$ & $37^{+10}_{-12}$ & B3/5(V) & 1978MSS...C02....0H & d \\
114 911 & * eta Mus & $793^{+309}_{-585}$ & $37^{+9}_{-23}$ & B7III+B7III & 1970MmRAS..72..233H & d \\
112 092 & * mu.01 Cru & $657^{+249}_{-480}$ & $35^{+11}_{-13}$ & B2IV & 1975A\&AS...19...91L & d \\
79 447 & * i Car & $961^{+696}_{-596}$ & $36^{+10}_{-11}$ & B3V & 2006MNRAS.371..252L & d \\
108 483 & * sig Cen & $971^{+974}_{-692}$ & $33^{+9}_{-12}$ & B3V & 1969MNRAS.144...31B  & d \\
120 307 & * nu. Cen & $753^{+821}_{-501}$ & $34^{+10}_{-11}$ & B2V & 1976A\&AS...26..241C & d \\
121 743 & * phi Cen & $1016^{+1147}_{-482}$ & $36^{+10}_{-9}$ & B2IV & 1969ApJ...157..313H  & d \\
129 116 & * b Cen & $1063^{+846}_{-561}$ & $37^{+9}_{-10}$ & B2.5V & 1968ApJ...151.1043S & d \\
136 298 & * del Lup & $675^{+515}_{-422}$ & $34^{+11}_{-10}$ & B1.5IV & 1969ApJ...157..313H & d \\
144 470 & * ome Sco & $820^{+1027}_{-360}$ & $37^{+9}_{-9}$ & B1V & 2024A\&A...690A.176N & d \\
147 165 & * sig Sco & $241^{+48}_{-31}$ & $35^{+7}_{-8}$ & B1III+B1:V & 2021A\&A...646A..11M & d \\
HDS 1106 & * P Pup & $600^{+307}_{-112}$ & $33^{+7}_{-5}$ & B0III  & 1969ApJ...157..313H  & d \\
2905 & * kap Cas & $744^{+121}_{-137}$ & $38^{+8}_{-10}$ & BC0.7Ia & 2024A\&A...690A.176N & d \\
151 890 & * mu.01 Sco & $322^{+134}_{-66}$ & $32^{+6}_{-6}$ & B1V+B & 1975A\&AS...19...91L & d \\
24 398 & * zet Per & $377^{+89}_{-81}$ & $37^{+10}_{-11}$ & B1Ib & 2024A\&A...690A.176N & d \\
143 118 & * eta Lup & $688^{+734}_{-295}$ & $36^{+9}_{-9}$ & B2.5IV & 1969ApJ...157..313H & d \\
151 985 & * mu.02 Sco & $472^{+241}_{-261}$ & $33^{+12}_{-10}$ & B2IV & 1969ApJ...157..313H & d \\
72 127 & HD 72127  & $743^{+670}_{-117}$ & $32^{+10}_{-5}$ &B3III/IV  & 1978MSS...C02....0H  & d \\
122 451 & * bet Cen  & $124^{+12}_{-8}$ & $31^{+4}_{-6}$ & B1III & 1969ApJ...157..313H & d \\
37 128 & * eps Ori & $678^{+194}_{-190}$ & $34^{+5}_{-2}$ &  B0Ia & 2024A\&A...690A.176N & d \\
23 180 & * omi Per   & $327^{+10}_{-10}$ & $33^{+4}_{-6}$ & B1III  & 2020A\&A...641A..35S  & d \\
35 411 & * eta Ori  & $552^{+200}_{-219}$ & $36^{+10}_{-9}$ & B0V  & 2018MNRAS.474.5287A & d \\
 200120 & * f01 Cyg & $677^{+35}_{-259}$ &  $35_{-11}^{+11}$ &B1.5Vnne &  1968ApJS...17..371L&d \\
\bottomrule
\end{tabular}
\end{table*}

\bsp	
\label{lastpage}
\end{document}